\documentclass[a4paper,num-refs]{oup-contemporary-modified}

\journal{gigascience}

\usepackage{graphicx}
\usepackage{siunitx}
\usepackage{xcolor}
\usepackage{float}

\title{Trajectories, bifurcations and pseudotime in large clinical datasets: applications to myocardial infarction and diabetes data}

\author[1]{Sergey E. Golovenkin}
\author[2,3,4]{Jonathan Bac}
\author[2,3,4]{Alexander Chervov}
\author[5,6]{Evgeny M. Mirkes}
\author[1]{Yuliya V. Orlova}
\author[2,3,4]{Emmanuel Barillot}
\author[5,6]{Alexander N. Gorban}
\author[2,3,4]{Andrei Zinovyev}

\affil[1]{Krasnoyarsk State Medical University, 660022, Krasnoyarsk, Russia}
\affil[2]{Institut Curie, PSL Research University, F-75005 Paris, France}
\affil[3]{INSERM, U900, F-75005 Paris,France}
\affil[4]{CBIO-Centre for Computational Biology, Mines ParisTech, PSL Research University, 75006 Paris, France}
\affil[5]{University of Leicester, LE1 7RH, Leicester, UK}
\affil[6]{Lobachevsky University, 603000 Nizhny Novgorod, Russia}

\runningauthor{Golovenkin et al.}

\begin{document}

\begin{frontmatter}
\maketitle
\begin{abstract}
    Large observational clinical datasets become increasingly available for mining associations between various disease traits and administered therapy. These datasets can be considered as representations of the landscape of all possible disease conditions, in which a concrete pathology develops through a number of stereotypical routes, characterized by `points of no return' and `final states' (such as lethal or recovery states). Extracting this information directly from the data remains challenging, especially in the case of synchronic (with a short-term follow up) observations. Here we suggest a semi-supervised methodology for the analysis of large clinical datasets, characterized by mixed data types and missing values, through modeling the geometrical data structure as a bouquet of bifurcating clinical trajectories. The methodology is based on application of elastic principal graphs which can address simultaneously the tasks of dimensionality reduction, data visualization, clustering, feature selection and quantifying the geodesic distances (pseudotime) in partially ordered sequences of observations. The methodology allows positioning a patient on a particular clinical trajectory (pathological scenario) and characterizing the degree of progression along it with a qualitative estimate of the uncertainty of the prognosis. Overall, our pseudo-time quantification-based approach gives a possibility to apply the methods developed for dynamical disease phenotyping and illness trajectory analysis (diachronic data analysis) to synchronic observational data. We developed a tool $ClinTrajan$ for clinical trajectory analysis implemented in Python programming language. We test the  methodology in two large publicly available datasets: myocardial infarction complications and readmission of diabetic patients data. 
    
\end{abstract}

\end{frontmatter}

\begin{keypoints*}
\begin{itemize}
\item Large-scale observational clinical datasets represent landscapes of the variety of disease states 
\item Diachronic clinical trajectories can be approximated from the multi-dimensional geometry of synchronic data and used for disease dynamical phenotyping
\item ClinTrajan: Python package for finding and analyzing clinical trajectories using elastic principal graph method
\end{itemize}
\end{keypoints*}


\section{Background}
\label{sec:background}

Large observational datasets are becoming increasingly available, reflecting physiological state of observed individuals, their lifestyles, exposure to environmental factors, received treatments and passed medical exams. From the big data point of view, each person's life can be represented as a trajectory in a multidimensional space of qualitative or quantitative traits. Simultaneous analysis of a large number of such trajectories can reveal the most informative features whose dynamics is correlated with trajectory clusters, associations between various factors and, potentially, the ``points of no return", i.e. bifurcations representing important fate decisions. 

The most important applications of such a framework are medical. The notion of ``disease trajectory'' as a person's trajectory in the data space of various diagnoses (diseases quantified by their severity) accompanying the person's life has emerged recently and became available for the large-scale analyses in certain contexts \cite{Jensen2014, Westergaard2019}. For example, a dataset containing an electronic health registry collecting during 15 years and covering the whole population of Denmark, with 6.2 million individuals, have been analyzed with an objective to determine previously unreported disease co-morbidities \cite{Jensen2014}. An ambitious `Data Health Hub' (\url{https://www.health-data-hub.fr/}) project has been recently launched in France with the aim to make available for machine learning-based analysis the collection of several decades-long population-wide anonymized health insurance records \cite{Moulis2015}. A formal review and meta-analysis of scientific texts using the concept of patient trajectory (or clinical pathway) based on disease management and care but also considering medico-economic aspects with a focus on myocardial infarction has been recently published in \cite{Pinaire2017}.

Dynamical phenotyping is the conceptual paradigm underlying such studies which can be applied at organismal and cellular scales \cite{Albers2014, Ruderman2017, Wang2017}. It states that distinguishing various dynamical types of progression of a disease or a cellular program is more informative than classifying biological system states at any fixed moment of time, because the type of dynamics is more closely related to the underlying hidden mechanism. From the machine learning point of view, this dictates different choices of methods, with clustering more adapted to the synchronic (snapshot) data \cite{Xu2008} while more specific methods for trajectory analysis are needed in the case of diachronic (having important temporal aspect) data \cite{Jung2008, Nagin2010, Rizopoulos2011, Schulam2015, Schulam2016}. The dynamical phenotyping paradigm and accompanying data mining methodologies become even more important with wider introduction of various types of continuous health monitoring devices and apps \cite{Banaee2013}.

However, examples of massive comprehensive and life-long longitudinal clinical data are still rare. Most of the existing clinical datasets correspond to relatively short periods of patients' stays inside hospitals, or during their treatment for a particular disease. In this sense, clinical datasets frequently represent detailed but rather ``static snapshot'' than the dynamical picture of the individuals' states. Nevertheless, one can hypothsize that such a snapshot, if sufficiently large, can sample the whole landscape of possible clinical states with certain routes and branches corresponding to some averaged illness trajectories. Then each patient can be thought to occupy a particular position along such a trajectory, where those patients following the same trajectory can be ranked accordingly to the progression along it from the hypothetical least heavy state towards some extreme state.

This situation is reminiscent of some recent studies of molecular mechanisms of several highly dynamical biological processes such as development or differentiation, at single cell level. Indeed, profiling a snapshot of a cell population can capture individual cells in the variety of different states (e.g., map their progression through the cell cycle phases). This allows reconstructing cellular trajectories through sampling the dynamics of the underlying phenomenon without necessity to follow each individual cell in time \cite{Chen2019, Saelens2019}. In this field, a plethora of machine learning-based methods have been recently suggested in order to capture the cellular trajectories and quantify progression along them in terms of {\it pseudo-time}, representing the total number of molecular changes in the genome-wide profiles of individual cells \cite{Saelens2019} rather than physical time. Many of these methods are able to detect branching trajectories, where the branches can represent important bifurcations (or, cell fate decisions) in the dynamical molecular processes underlying differentiation of developmental programs. 

The aim of the present study is to suggest and test a computational methodology for extracting clinical trajectories from sufficiently large synchronic clinical datasets. Clinical trajectory is a clinically relevant sequence of ordered patient phenotypes representing consecutive states of a developing disease and leading to some final state (i.e., a lethal outcome). Importantly, in our approach we do not assume that these are the same patient's states, even if this can be so in the case when there exist some longitudinal observations. Each clinical trajectory can be characterized by its proper pseudotime which allows one to quantitatively characterize the degree of progression along the trajectory. Each clinical variable can be analyzed as a function of pseudotime conditioned on a given clinical trajectory. We also assume that clinical trajectories can be characterized by branches (bifurcations), representing important decisive moments in the course of a disease. 

An important methodological difference between the previously developed methodology of cell trajectory analysis in omics datasets, where the majority of the variables can be considered continuous and of similar nature  (e.g., gene expression levels), the clinical datasets possess certain specifics which must be taken into account. Typical real-life clinical data are characterized by the following features: a) they contain mixed data types (continuous, binary, ordinal, categorical variables, censored data); b) they typically contain missing values with non-uniform missingness pattern across the data matrix; c) they do not have a uniquely defined labeling (part of the clinical variables can be used to define clinical groups, but this can be done in several meaningful ways). This means that an important integral part of the methodology should be procedures for quantifying and imputing missing values in mixed type datasets making them amenable for further application of machine learning methods. The last feature (c) suggests that unsupervised or semi-supervised methodology might play more important and insightful role here than purely supervised methods.

We develop a methodology of clinical data analysis, based on modeling the multi-dimensional geometry of a clinical dataset as a ``bouquet" of diverging clinical trajectories, starting from one or several quasi-normal (least severe) clinical states. As a concrete approach we exploit the methodology of elastic principal trees (EPT), which is a non-linear generalization of Principal Component Analysis (PCA). Principal tree is a set of principal curves assembled in a tree-like structure, characterized by branching topology \cite{Gorban2007Topological, Gorban2008Principal}. Principal trees can be constructed using ElPiGraph computational tool, which has been previously exploited in determining branching trajectories in various genomics datasets (in particular, in single cell omics data) \cite{Albergante2020, Chen2019, Parra2019}. As an unsupervised machine learning method, estimating elastic principal graphs solves several tasks simultaneously, namely dimensionality reduction, data visualization, partitioning the data by the non-branching graph segments (analogous to clustering) and quantifying robust geodesic distances (pseudo-times) from one data point to another along the reconstructed principal graph. Unlike many other methods relying on heuristics for guessing the optimal graph topology (e.g., a tree) such as Minimal Spanning Tree (MST), elastic principal graph method optimizes the graph structure via application of topological grammars and gradient descent-like optimization in the discrete space of achievable graph structures (e.g., all possible tree-like graphs)\cite{Albergante2020}.

The suggested method is implemented as Python package, $ClinTrajan$, which can be easily used in the analysis of clinical datasets. We provide several reproducible Jupyter notebooks illustrating the different steps of the methodology. The figures in this paper are directly copied from these notebooks. The methodology proved to be scalable to the datasets containing hundreds of thousands of clinical observations, using an ordinary laptop, and can be scaled up further for even larger datasets.

\section{Data Description}

In this study we apply the suggested methodology to two publicly available clinical datasets, one of moderate size (1700 patients) and one of relatively large size (>100,000 patients).

\subsection{Complications of myocardial infarction database}

Myocardial infarction (MI) is one of the most dangerous diseases. The wide spread of this disease over the past half century has made it one of the most acute problems of modern medicine. The incidence of myocardial infarction (MI) remains high in all countries. This is especially true of the urban population of highly developed countries, exposed to the chronic effects of stress factors, irregular and not always balanced nutrition. In the United States annually, more than one million people become ill with myocardial infarction \cite{Marso1999}.

The course of the disease in patients with MI is diverse. MI can occur without complications or with complications that do not worsen the long-term prognosis. At the same time, about half of patients in the acute and subacute periods have complications leading to a worsening of the course of the disease and even death. Even an expert can not always foresee the development of these complications. In this regard, predicting the complications of myocardial infarction in order to timely carry out the necessary preventive measures seems to be an important task.

The database analyzed here was collected in the Krasnoyarsk Interdistrict Clinical Hospital (Russia) in 1992-1995 years, but has only recently been deposited to the public domain. The original database and its description can be downloaded from \cite{Golovenkin2020}. It contains information about 1700 patients characterized by 111 features describing the clinical phenotypes and 12 features representing possible complications of the myocardial infarction disease (123 features in total). Previously, the dataset was a subject of machine learning method applications, including convolutional neural networks \cite{Gorban1995MultiNeuron} and dimensionality reduction methods \cite{Zinovyev2001}. We believe that introducing this dataset, which exemplifies the specificity and difficulties of analysing the real-life clinical data, to the big data and machine learning research community should contribute to developing better treatment and subtyping strategies in cardiology and in clinical research in general \cite{Potluri2016}.

The detailed description of the variable names with associated descriptive statistics is provided in the dataset description available online \cite{Golovenkin2020}. Here we provide Table~\ref{tab:infarctus_variable_names} with the meaning of those variables which appear in the figures of the manuscript.

\begin{table*}[bt!]
\caption{Names of selected variables from the myocardial infarction complication dataset}\label{tab:infarctus_variable_names}
\begin{tabularx}{\linewidth}{l L l L}
\toprule
{Variable name} & {Meaning} & {Variable name} & {Meaning} \\
\midrule
\multicolumn{4}{c}{General input values}\\
AGE & Age & DLIT\_AG & Duration of arterial hypertension \\
ant\_im & Presence of an anterior myocardial infarction (left ventricular) & FK\_STENOK & Functional class of angina pectoris in the last year \\
GIPER\_NA & Increase of sodium in serum (more than 150 mmol/L) & IBS\_POST & Coronary heart disease in recent weeks, days before the admission time\\
inf\_im & Presence of an inferior myocardial infarction (left ventricular) & lat\_im & Presence of a lateral myocardial infarction (left ventricular) \\
K\_BLOOD & Serum potassium content (mmol/L) & L\_BLOOD & White blood cell count (billions per liter)\\

NA\_BLOOD & Serum sodium content (mmol/L) & post\_im & Presence of a posterior myocardial infarction \\
NA\_R\_1\_n & Use of opioid drugs in the ICU in the first hours of the hospital period & NA\_R\_2\_n & Use of opioid drugs in the ICU in the second day of the hospital period\\
NA\_R\_3\_n & Use of opioid drugs in the ICU in the third day of the hospital period & NOT\_NA\_1\_n & Use of NSAIDs in the ICU in the first hours of the hospital period\\
NOT\_NA\_2\_n & Use of NSAIDs in the ICU in the second day of the hospital period & NOT\_NA\_3\_n & Use of NSAIDs in the ICU in the third day of the hospital period \\
R\_AB\_1\_n & Relapse of the pain in the first hours of the hospital period & R\_AB\_2\_n & Relapse of the pain in the second day of the hospital period\\
R\_AB\_3\_n & Relapse of the pain in the third day of the hospital period & TIME\_B\_S & Time elapsed from the beginning of the attack of CHD to the hospital\\

\multicolumn{4}{c}{Inputs from anamnesis}\\
nr\_03 & Paroxysms of atrial fibrillation & nr\_04 & A persistent form of atrial fibrillation \\
nr\_11 & Observing of arrhythmia & np\_10 & Complete RBBB \\
STENOK\_AN & Exertional angina pectoris & zab\_leg\_02 & Obstructive chronic bronchitis\\
zab\_leg\_03 & Bronchial asthma & zab\_leg\_06 & Pulmonary tuberculosis\\
ZSN\_A & Presence of chronic heart failure &&\\

\multicolumn{4}{c}{Inputs for the time of admission to hospital}\\
n\_p\_ecg\_p\_06 & Third-degree AV block on ECG & n\_p\_ecg\_p\_08 & LBBB (posterior branch) on ECG \\
n\_p\_ecg\_p\_12 & Complete RBBB on ECG & n\_r\_ecg\_p\_05 & Paroxysms of atrial fibrillation on ECG \\
n\_r\_ecg\_p\_06 & Persistent form of atrial fibrillation on ECG & n\_r\_ecg\_p\_08 & Paroxysms of supraventricular tachycardia on ECG \\
ritm\_ecg\_p\_01 & Sinus ECG rhythm (HR between 60 and 90) & ritm\_ecg\_p\_02 & Atrial fibrillation in ECG rhythm\\
ritm\_ecg\_p\_04 & Atrial ECG rhythm & ritm\_ecg\_p\_06 & Idioventricular  ECG rhythm\\
ritm\_ecg\_p\_07 & Sinus ECG rhythm (HR above 90) & SVT\_POST & Paroxysms of supraventricular tachycardia\\

\multicolumn{4}{c}{Inputs for the time of admission to intensive care unit}\\
D\_AD\_ORIT & Diastolic blood pressure (mmHg) & S\_AD\_ORIT & Systolic blood pressure (mmHg) \\
FIB\_G\_POST & Ventricular fibrillation & K\_SH\_POST & Cardiogenic shock \\
MP\_TP\_POST & Paroxysms of atrial fibrillation & O\_L\_POST & Pulmonary edema \\

\multicolumn{4}{c}{Complications}\\
FIBR\_PREDS   &    Atrial fibrillation & PREDS\_TAH & Supraventricular tachycardia \\
JELUD\_TAH   &  Ventricular tachycardia & FIBR\_JELUD & Ventricular fibrillation \\
DRESSLER & Dressler syndrome & ZSN & Chronic heart failure \\
OTEK\_LANC & Pulmonary edema & P\_IM\_STEN & Post-infarction angina \\
REC\_IM & Relapse of the myocardial infarction & A\_V\_BLOK & Third-degree AV block \\
RAZRIV & Myocardial rupture &&\\

\multicolumn{4}{c}{Cause of death}\\
LET\_IS=0 & Survive & LET\_IS>0 & Death with cause from 1 to 7 \\
LET\_IS\_0 & Survive & LET\_IS\_1 & Cardiogenic shock \\
LET\_IS\_2 & Pulmonary edema & LET\_IS\_3 & Myocardial rupture\\
LET\_IS\_4 & Progress of congestive heart failure & LET\_IS\_5 & Thromboembolism\\
LET\_IS\_6 & Asystole & LET\_IS\_7 & Ventricular fibrillation\\

\bottomrule
\end{tabularx}

\end{table*}

\subsection{Diabetes readmission data set}

Together with myocardial infarction, various diabetes-related clinical states such as hyperglycemia are widely spread in the modern population. The management of hyperglycemia in the hospitalized patients has a significant bearing on outcome, in terms of both morbidity and mortality \cite{Strack2014}. An assembly and analysis of a large clinical database was undertaken to examine historical
patterns of diabetes care in patients admitted to a US hospital and to inform future directions which might lead to improvements in patient safety \cite{Strack2014}. In particular, the use of HbA1c as a marker of attention to diabetes care in a large number of individuals identified as having a diagnosis of diabetes mellitus was analyzed. A focus was on the readmission probability of a patient after leaving the hospital and its dependency on other clinical features that can be collected during hospitalization.

The dataset represents 10 years (1999-2008) of clinical care at 130 US hospitals and integrated delivery networks. It includes over 50 features representing patient and hospital outcomes. The dataset can be downloaded from UCI repository at \url{https://archive.ics.uci.edu/ml/datasets/diabetes+130-us+hospitals+for+years+1999-2008} or from Kaggle at \url{https://www.kaggle.com/brandao/diabetes}. The data  contains more than 100,000 hospitalization cases with patients suffering from diabetis characterized by 55 attributes.

\section{Analyses}

\subsection{ClinTrajan package for trajectory inference in large clinical datasets}

We suggest computational methodology of constructing principal trees in order to extract clinical trajectories from large-scale clinical datasets which takes into account their specificity. The following steps of the analysis have been implemented:

\begin{itemize}
\item Univariate and multi-variate quantification of nominal variables, including an original implementation of the optimal scaling procedure for ordinal values
\item Several methods for missing values imputation including two original implementations of SVD-based imputers
\item Constructing principal tree for a quantified clinical dataset
\item Partitioning the data accordingly to the non-branching segments of the principal tree (analogue of clustering) and associating the segments to clinical variables
\item Extracting clinical trajectories and associating the trajectories to clinical variables
\item Visualization of clinical variables using principal trees and metro map data layouts \cite{Gorban2008Beyond}
\item Pseudotime plots of clinical variables along clinical trajectories, visualization of their bifurcations
\end{itemize}

\subsection{Myocardial infarction complications case study}

\subsubsection{Quantification of nominal values and imputation of missing data values}

As a first step of pre-processing, 7 variables have been removed from the initial myocardial infarction complication data table, as containing more than 30\% of missing values. Afterwards, 126 records have been removed as containing more than 20\% of missing values. After this step, the data table contained 2.5\% of missing values with 533 rows (34\% of all clinical cases) having no missing values.

After the missing value filtering step, the data table of myocardial infarction complications contained 84 binary, 9 continuous numerical, 22 ordinal and 1 categorical variables. Large number of ordinal variables requires careful quantification of them (see Methods), which is not trivial given the large number of rows with missing values.

We considered that the small number of continuous numerical variables is not enough to apply the methodology of Categorical Principal Component Analysis (CatPCA) \cite{Casacci2015}. Therefore, for all ordinal and binary variables we first applied the univariate quantification following the approach described in the `Methods' section.  This quantification allowed applying `SVDComplete' imputation method for imputing the missing values, as described in `Methods'. After all missing values have been imputed, we could apply the optimal scaling approach for ordinal values, optimizing the pairwise correlations between them and between ordinal and continuous numerical variables. The 22 ordinal variables quantified in this way were further used for forming the data space. In addition, all variables were converted to z-scores. 

\subsubsection{Constructing elastic principal tree}

The initial data space was formed by 123 variables. 
We evaluated the global intrinsic dimension of the dataset using several methods implemented in \url{https://github.com/j-bac/scikit-dimension}, and found that the majority of non-linear methods estimate the intrinsic dimension in the range 10-15 while linear methods based on PCA gives much larger intrinsic dimension values (see Supplementary Figure 1). We compared the estimations of intrinsic dimensions with and without complication variables and found them to be similar, which indicates that there exists a certain level of dependency between the complication variables and the rest of the clinical variables. We also observed that the screen plot for this dataset is characterized by elbow approximately at $n=12$. As a result of this analysis, for further inference of the principal tree, we projected the dataset into the space of the first 12 principal components.

\begin{figure*}[bt!] 
\centering
\includegraphics[width=\linewidth]{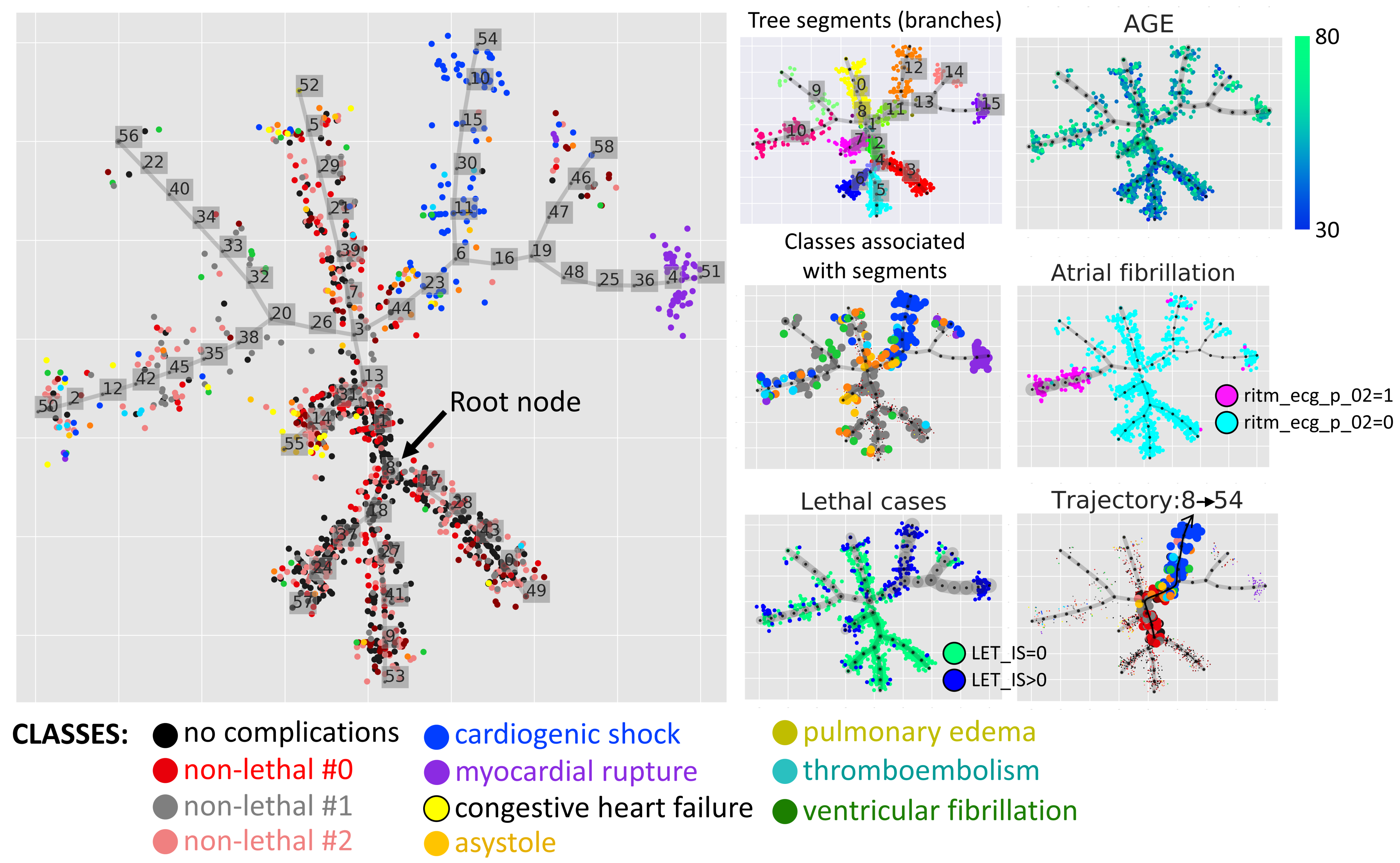}
\caption{Principal tree recapitulating the multidimensional structure of the myocardial infarction complications dataset. Distribution of classes along the tree is visualized in the large panel. Various modes of data visualization are shown in the small panels. `Tree segments (branches)' shows partitioning (clustering) of data points accordingly to the linear fragments connecting branching points and/or leaf nodes (called `non-branching segments' in this work). `AGE' is an example of a continuous variable visualization using color gradient of data points. `Classes associated with segments' shows data points of only those classes which have statistical associations with one or more segments. `Artrial fibrillation' and `Lethal cases' show visualization of a binary variable, with edge width reflecting the trend (in case of `Lethal cases' it can be interpreted as lethality risk estimate). `Trajectory 8$\rightarrow$ 54' a subset of data points, colored accordingly to their classes and belonging to one particular clinical trajectory, having the node with the least risk of complications as the root node (node 8) and the highest risk of cardiogenic shock as its final state (node 54). }\label{fig:InfarctusPT}
\end{figure*}

The elastic principal tree was computed using ElPiGraph Python implementation as documented in the Jupyter notebook provided at \url{https://github.com/auranic/ClinTrajan} and in the Section
`Method of Elastic Principal Graphs (ElPiGraph)'
of this manuscript. The principal tree explained 52.4\% of total variance in contrast to the first two principal components that explained 25.9\% and first five PCs explaining 54\%. The obtained principal tree (shown in Figure~\ref{fig:InfarctusPT}) was used to provide a 2D layout of the dataset which can be used for visualization of various clinical variables and the results of analyses. Globally, the principal tree defined three terminal non-branching segments populated with non-lethal clinical cases (indicated as \#3,\#5,\#6 in Figure~\ref{fig:InfarctusPT}, panel `Tree segments (branches)') and associated with younger patients (Figure~\ref{fig:InfarctusPT}, panel `AGE'). Other terminal segments (\#0,\#7,\#9,\#10,\#12,\#14,\#15) were characterized by various risks of lethality (Figure~\ref{fig:InfarctusPT}, panel `Lethal cases'), with two terminal segments \#12 and \#15 being strongly enriched with lethal cases, caused by cardiogenic shock and myocardial rupture correspondingly. 

Each node of the principal tree is connected with a subset of data points. We performed the enrichment analysis, based on application of independence $\chi^2$ test in order to determine the node which is the most strongly associated to `no complication' class (black points in Figure~\ref{fig:InfarctusPT}). The position of this node (\#8) is indicated as `Root node' in Figure~\ref{fig:InfarctusPT}, main panel.

\subsubsection{Assigning data point classes}

As we've mentioned above, the classes of the clinical observations can be usually defined by selecting a subset of clinical variables which represent some final read-outs of a patient state. Thus, in the myocardial infarction complications dataset, 12 clinical variables report the complications, 11 of them represent binary variables and 1 categorical variable LET\_IS, whose value is 0 if there is no lethal outcome. Otherwise, LET\_IS can take one of the 7 nominal values representing the death cause. Following the methodology suggested in this study, the LET\_IS variable is first made a subject of dummy coding, introducing 7 binary features. The resulting 18 binary variables were characterized by 158 unique combinations of 0/1 values, which looked too many to define one class per such unique combination.

\begin{figure}[bt!] 
\centering
\includegraphics[width=\linewidth]{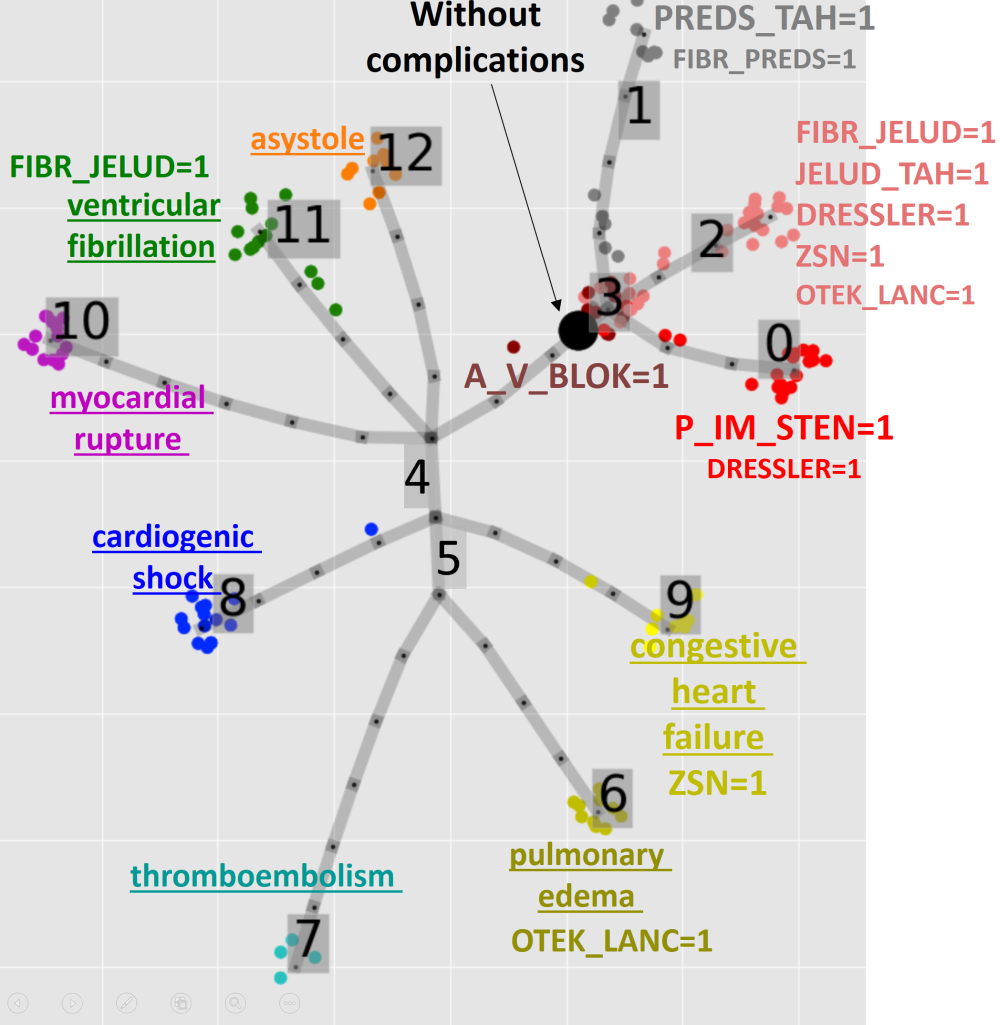}
\caption{Defining classes of myocardial infarction complications, using principal tree-based clustering. The labeling marks either the cause of death (underlined, for lethal outcome classes) or a set of complication variables strongly overrepresented in the cluster, accordingly to the $\chi^2$ test of independence; the size of the label reflects the significance of over-representation (the larger the label the more significant is the deviation from independence). The meaning of the complication variables here are:
FIBR\_PREDS - atrial fibrillation, PREDS\_TAH - supraventricular tachycardia,
JELUD\_TAH - ventricular tachycardia, FIBR\_JELUD - ventricular fibrillation, A\_V\_BLOK - third-degree AV block, OTEK\_LANC - pulmonary edema, DRESSLER - Dressler syndrome, ZSN - chronic heart failure, REC\_IM - relapse of the myocardial infarction, P\_IM\_STEN - post-infarction angina.
}\label{fig:InfarctusClasses}
\end{figure}

Therefore, it was decided to reduce the number of the distinct complication states to a more manageable number by clustering them. The table of 158 possible complications and 18 binary variables was analyzed by the method of elastic principal trees as described below and clustered into 11 clusters accordingly to the principal tree non-branching segments (see Figure~\ref{fig:InfarctusClasses}). 7 of these clusters contained lethal outcomes and clearly corresponded to particular death causes, which corresponds to non-zero values of LET\_IS variable. The non-lethal outcomes have been clustered into 4 classes (`0',`1',`2',`3'). Classes `1' and `2' appeared to be characterized by fibrillation and tachycardia, but differed in the types (`1' corresponded to the `atrium' fibrillation and tachycardia while `2' had a tendency to be characterized by the ventricle ones). Non-lethal class `0' was distinguished by `P\_IM\_STEN' (post-infarction angina), and the class `3' -- by the presence of diagnosed `A\_V\_BLOCK' (third-degree atrioventricular block).

Besides this clustering, a particular non-lethal state was distinguished characterized by zero values of all complication variables. We distinguished this class as a separate `no complications class'. In the complete dataset, it corresponded to 45\% of clinical records (denoted as black points in Figure~\ref{fig:InfarctusClasses} and Figure~\ref{fig:InfarctusPT}). In the rest of the analysis, all complication variables have been analyzed together with the clinical characteristics.

\subsubsection{Dataset partitioning (clustering) by principal tree non-branching segments}

The explicitly defined structure of the computed elastic principal tree allows partitioning the dataset accordingly to the projection of the data points on various internal and terminal segments as described in the `Methods' section and shown by color in Figure~\ref{fig:InfarctusPT}, panel `Tree segments (branches)'. Such partitioning can play a role of clustering with the advantage of that the tree segments can recover non-spherical and non-linear data clusters. In addition, the data clusters, defined in such a way, are connected in a tree-like configuration, with junctions corresponding to the branching points which can correspond to `points of no return' in the state of the patients.

\begin{figure*}[bt!] 
\centering
\includegraphics[width=\linewidth]{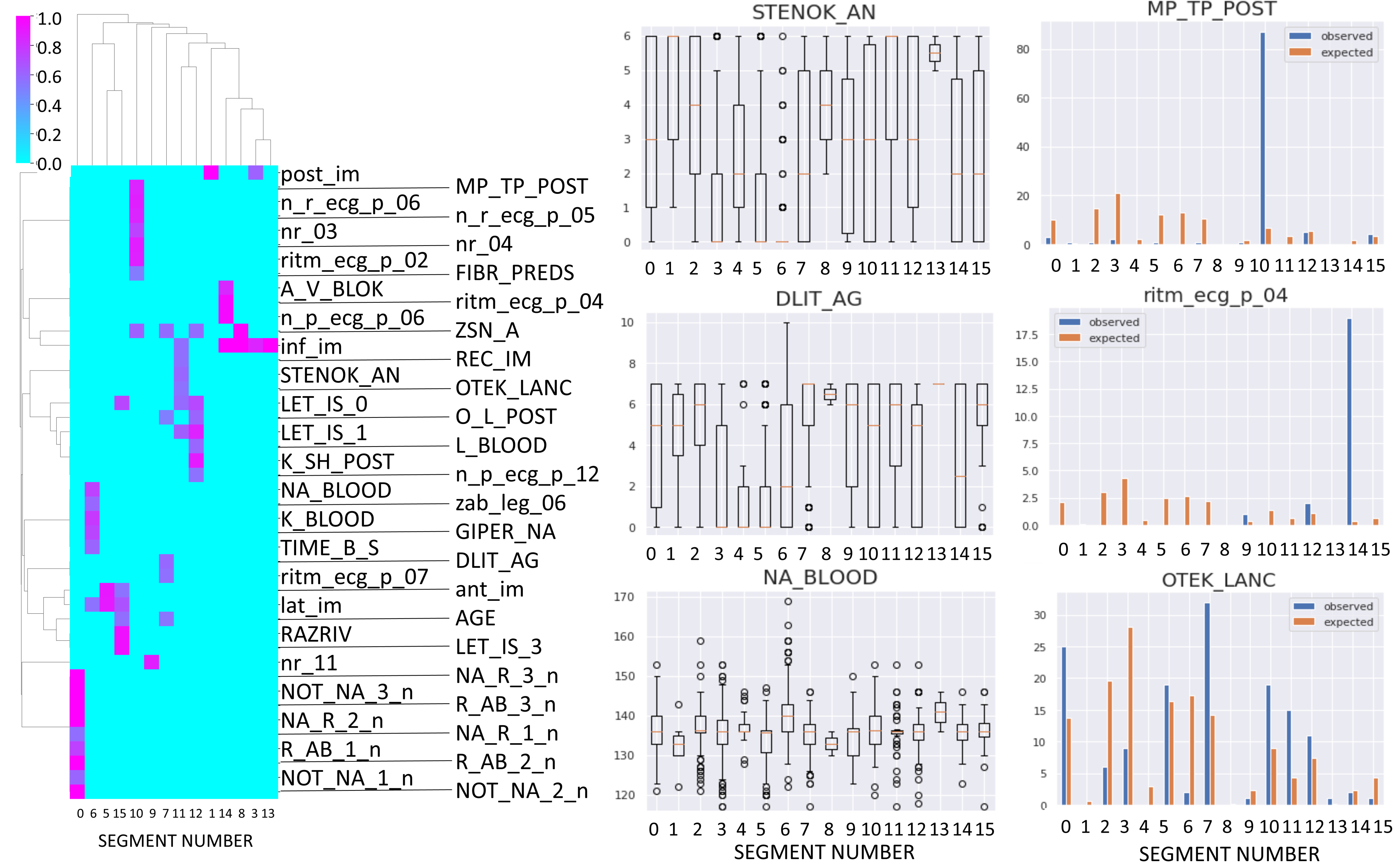}
\caption{Association of principal tree segments (as shown in Figure~\ref{fig:InfarctusPT}) with data variables. On the left, a hierarchical clustering dendrogram of association scores is shown. On the right, three examples of strong associations with continuous/ordinal and binary variables are shown. Here the  following variables have been shown: STENOK\_AN, Exertional angina pectoris in the anamnesis, DLIT\_AG, Duration of arterial hypertension (years), NA\_BLOOD, Serum sodium conten, mmol/L, MP\_TP\_POST, Paroxysms  of  atrial  fibrillation, ritm\_ecg\_p\_04, ECGrhythm at the time of admission to hospital–atrial, OTEC\_LANC, Pulmonary edema.  The meaning of other variable names is provided in the ``Data description" section. }\label{fig:InfarctusBranches}
\end{figure*}

Each non-branching segment in the tree can be associated by enrichment analysis either to a data class or to a variable. The points of the data classes which are associated to at least one tree segment are highlighted by size in Figure~\ref{fig:InfarctusPT}, panel `Classes associated with segments'. The results of enrichment analysis for all clinical variables are shown in Figure~\ref{fig:InfarctusBranches}. Briefly, we found that 44 clinical variables, including 8 complication variables, can be associated to at least one segment (Figure~\ref{fig:InfarctusBranches}) with reasonably high thresholds either for the deviation score (\ref{deviation_score}) or ANOVA linear model coefficient (provided that the results of chi-square or ANOVA tests are statistically significant). 

\subsubsection{Analysis of clinical trajectories and pseudotime}

The non-branching segments of the principal tree are connected into trajectories, from the root node of the tree corresponding to the least frequency of complications to one of the leaf nodes representing some extreme states of the disease (some of which are connected with increased risk of lethality). Internal tree segments are shared between several trajectories, while the terminal segments correspond to one single trajectory. Consequently, each data point can be associated to one or more trajectories. The position of the data point on a trajectory is quantified by the value of pseudo-time characterizing the intrinsic geodesic distance from the root node, measured in the units of the number of tree edges. The value of pseudo-time is continuous since a data point can be projected on a tree edge, in between two nodes.

If a data point (clinical observation) is attributed to several trajectories then it is characterized by the same pseudo-time value on each of them. This can be interpreted as the state of uncertainty from which several clinical scenarios can be developed in the further course of the disease, following one or several bifurcation points. Those clinical observations belonging to a single trajectory correspond to less uncertainty in the prognosis, with higher chances to end up in a terminal state.

\begin{figure*}[p] 
\centering
\includegraphics[width=16cm]{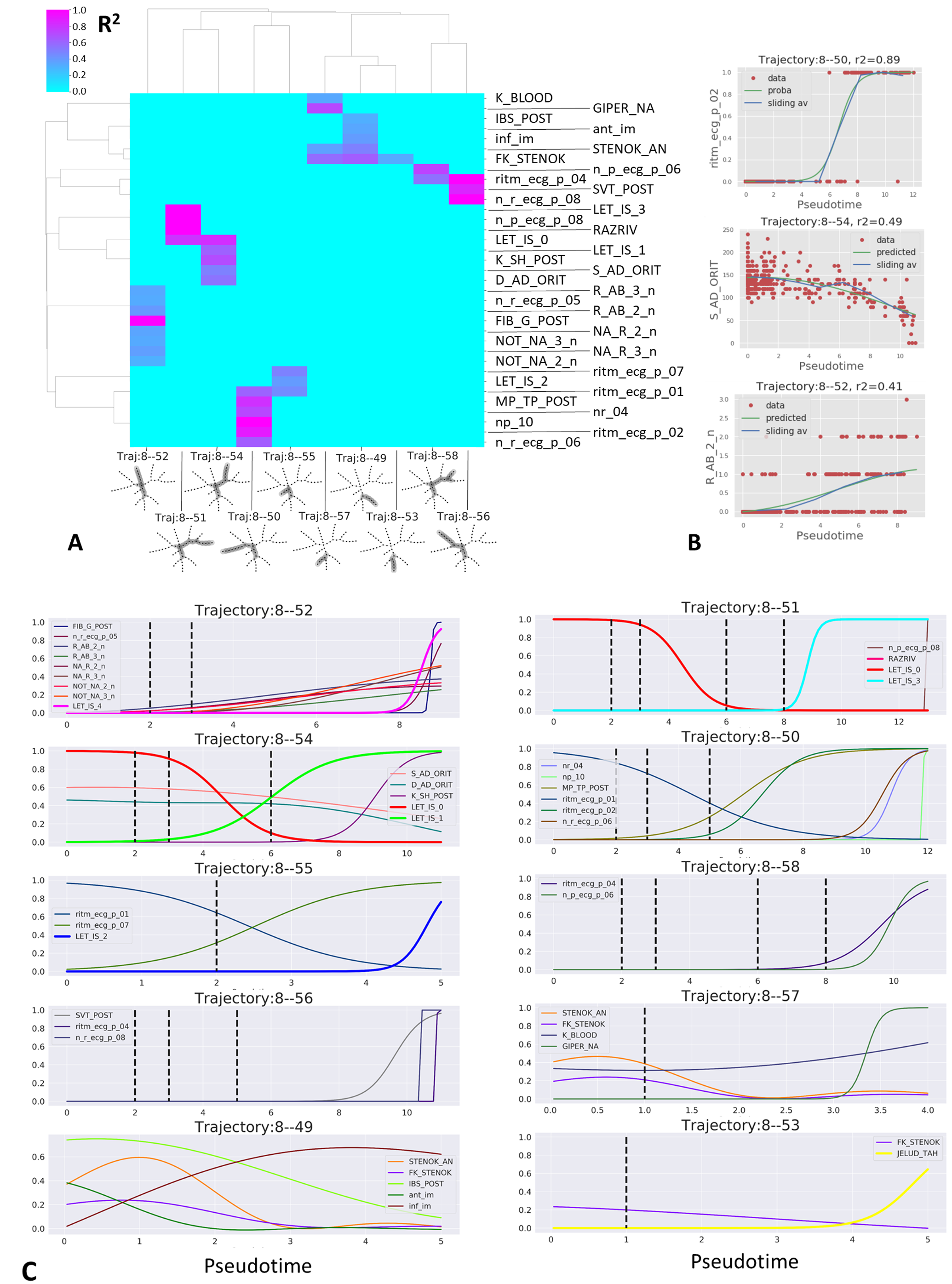}
\caption{Clinical trajectory analysis of the myocardial infarction complications dataset. A) Visualization of $R^2$ values for the regression between clinical variables and the pseudo-time along 10 clinical trajectories. B) Examples of regression analysis for a binary (logistic regression), continuous and ordinal (Gaussian kernel regression) clinical variables. C) Pseudotime plots for clinical variables selected by regression analysis. For binary variables, the probability inferred by logistic regression is shown. For ordinal and continuous variables, non-linear regression line is shown.  Complication  variables associated to the clinical trajectories are shown with thick lines (for example, LET\_IS\_0 represents the survival probability.) Vertical dashed lines indicate the positions of tree branching points along pseudo-time. The abscissa in the pseudotime plots corresponds to the variable value scaled to unity for the total variable  amplitude. The meaning of the variable names is provided in the ``Data description" section. }\label{fig:InfarctusTrajectory}
\end{figure*}

In order to determine the factors affecting the choices between alternative clinical trajectories, it is necessary to associate clinical variables to each of the trajectory and determine the trend of their changes along them. Mathematically this corresponds to the solving the regression problem connecting a clinical variable and the observation pseudo-time. Using this approach we identified 35 variables associated with pseudo-time with $R^2>0.3$ for at least one trajectory (Figure~\ref{fig:InfarctusTrajectory},A,B). The pseudo-dynamics of these variables is shown in Figure~\ref{fig:InfarctusTrajectory},C. This analysis allows one to conclude on the sequence of clinical variable changes leading to various complications. Thus, four trajectories $8\rightarrow 52, 51, 54, 55$ are associated with increasing risks of four distinct lethal outcomes (progress of congestive heart failure, myocardial rupture, cardiogenic shock and pulmonary edema respectively). Three trajectories ($8\rightarrow 49, 53, 57$) correspond to mild course of the disease associated with younger patients with the risk of ventricular tachycardia increasing along the trajectory $8\rightarrow 53$. 

\begin{figure*}[bt!] 
\centering
\includegraphics[width=\linewidth]{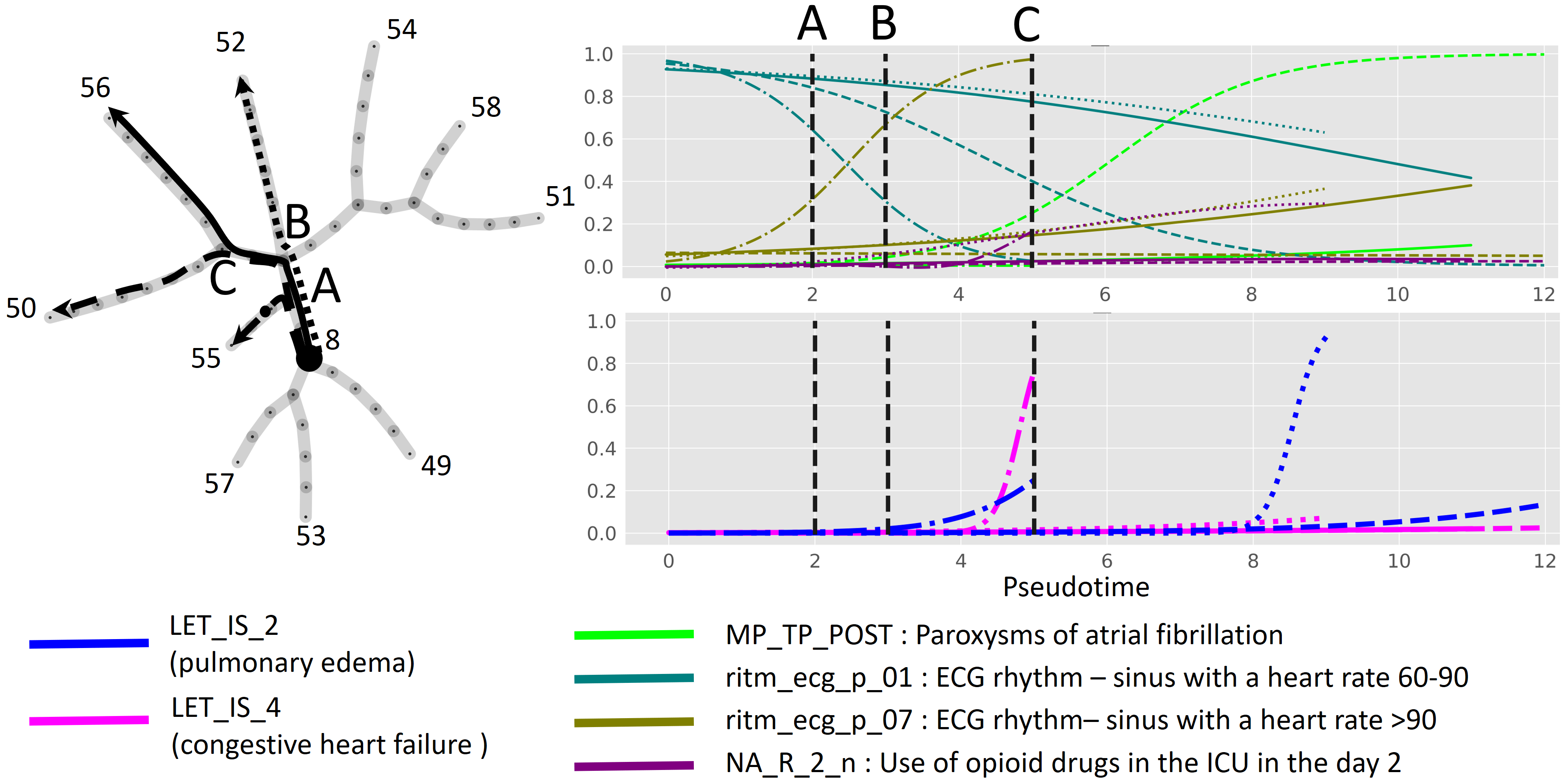}
\caption{Example of bifurcating clinical trajectories. Four clinical trajectories out of ten are depicted for myocardial infarction complications data. The trajectories all share the internal segment 8-A and diverge at nodes A, B, C. Four selected binary clinical variables and two lethal outcome variables are shown as functions of four pseudotime measurements, one per trajectory. The abscissa in the pseudotime plots corresponds to the variable value scaled to unity for the total variable  amplitude.}\label{fig:InfarctusTrajBifurcations}
\end{figure*}

In order to illustrate the picture of decreasing uncertainty while the disease progress along clinical trajectories, we focused on four trajectories $8\rightarrow 55, 50, 56, 52$ sharing one or several internal tree segments. We selected several clinical variables and two lethal outcome variables associated with pseudo-time along these trajectories and showed them all in one plot (Figure~\ref{fig:InfarctusTrajBifurcations}). One can see that the pseudo-dynamics of some clinical variables estimated by logistic regression as the probability of value `1', gradually diverge at the branching points of the principal tree. 

The trajectory $8\rightarrow 55$ is characterized by increasing sinus tachycardia after the bifurcation point A and increasing risk of congestive heart failure and to lesser extent the pulmonary edema. The trajectory $8\rightarrow 50$ is characterized by the absence of sinus tachycardia with gradual decline in the variable `ECG rhythm - sinus with a heart rate 60-90' after the point B, and, after the bifurcation point C, rapid increase of the probability of paroxysms of atrial fibrillation. The prognosis along this trajectory is relatively favorable as well as on the clinical trajectory $8\rightarrow 56$, which is characterized by slow and incomplete decrease of the probability of `ECG rhythm - sinus with a heart rate 60-90' after the point C. 

The trajectory $8\rightarrow 52$ is characterized by high risk of pulmonary edema and gradual increase of sinus tachycardia. One of the distinguishing features of this trajectory is increased use of opioid and antiinflammatory drugs in the intensive care unit at days 2 and 3 after the admission to the hospital, which is in turn connected to the pain relapse (R\_AB\_2\_n variable). 

\subsubsection{Predicting survival and lethal risk factors}

Each clinical trajectory extracted from the analysis of synchronic clinical data is interpreted as a possible ordered sequence of states from the least heavy condition to the extreme final point of the trajectory. Assuming that for a given patient state all the downstream points on the clinical trajectory represent possible future states of the patient, we can make a prediction of possible clinical risks connected with moving along this trajectory. In particular, this can be used for estimating lethal risks, is such events are recorded in the clinical dataset. In order to evaluate the risks of a clinical event in the future, well-developed methodology of survival analysis can be used, but using the pseudotime value instead of the real time value. We will call such analysis the pseudotime survival analysis. 

The pseudotime quantified along different clinical trajectories might be incomparable in terms of the physical time. Therefore, the pseudotime survival analysis should be performed for each trajectory individually, even if the estimated risks can be visualized together using the common pseudotime axis. 

\begin{figure*}[bt!] 
\centering
\includegraphics[width=\linewidth]{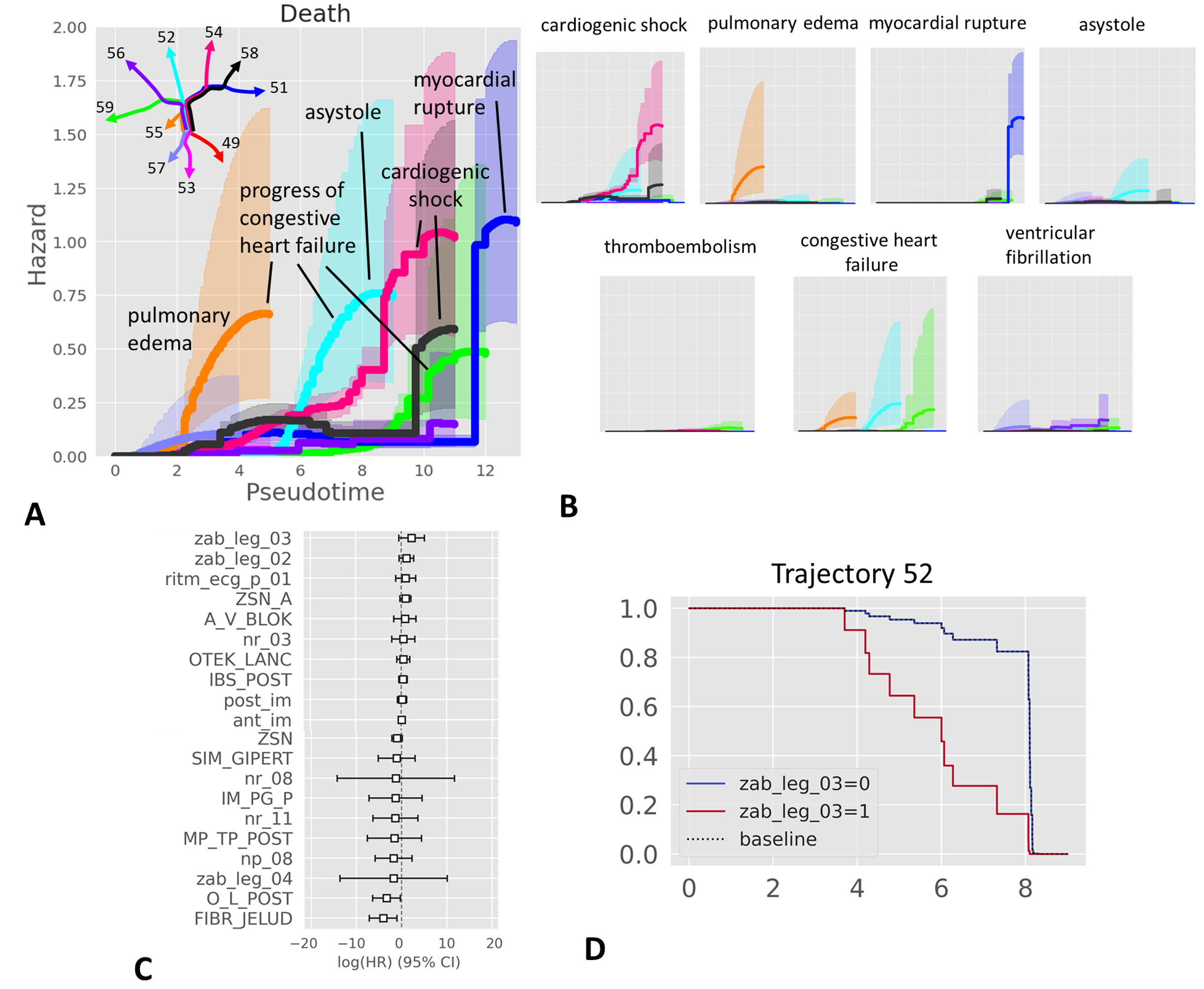}
\caption{Pseudotime survival analysis and determination of risk factors along clinical trajectories. A) Visualizing total hazards of death from myocardial infaction complications together with uncertainty of their estimate along different pseudo-time trajectories. The trajectories of the principal tree are denoted by different colors, corresponding to the color of the hazard plot. The dominant contribution of the death cause to the total hazard is annotated by a label. B) Hazards of individual causes of deaths along various trajectories. The axes scale of each small plot here is identical to the plot shown in A). C) Example of survival regression for the data points along the trajectory ending with the node 52 (denoted as light blue in A)). Only 10 most significant positive and 10 most significant negative survival regression coefficients are shown. D) The effect of the top positive survival regression coefficient (zab\_leg\_03, meaning presence of bronchial asthma in the anamnesis) leads to different survival functions between two patient groups. Thus, presence of asthma in the anamnesis (zab\_leg\_03=1) worsen the survival along this particular trajectory associated to the risk of congestive heart failure and asystole (cardiac arrest). }\label{fig:InfarctusSurvivalAnalysis}
\end{figure*}

We applied a non-parametric estimator of the cumulative hazard rate function (see Methods) in order to quantify lethal risks along ten identified clinical trajectories in the myocardial infarction complication dataset (Figure~\ref{fig:InfarctusSurvivalAnalysis},A). This analysis highlighted 6 out of 10 trajectories as characterized by elevated hazard rates of lethality, which is a quantification of the distribution of lethal cases on the principal tree shown before in Figure~\ref{fig:InfarctusPT}. The total lethality risk can be decomposed into the risks resulting from a particular death cause (one out of seven). Quantification of individual death cause risks is shown in Figure~\ref{fig:InfarctusSurvivalAnalysis},B. In this case, an event for the hazard function estimator is a particular death cause. As a result, the increased risk of total lethality can be attributed to one or several particular death causes (Figure~\ref{fig:InfarctusSurvivalAnalysis},A).

Using the same assumptions, different risk factors affecting the risks along different clinical trajectories can be evaluated using the standard methodology of survival regression. As an example, we computed the survival regression for the set of patients along the trajectory ending in the node '52', associated with increased risks of congestive heart failure and asystole (cardiac arrest). The clinical variable having the largest positive contribution to the regression was the presence of bronchial asthma in the anamnesis, suggesting that it can be an aggravating factor along this particular clinical trajectory (Figure~\ref{fig:InfarctusSurvivalAnalysis},C). Indeed, splitting this set of patients into two groups (having the asthma in anamnesis and not) shows differential survival as a function of pseudotime along this particular trajectory.

\subsection{Diabetes readmission case study}

\subsubsection{Clinical trajectories in large-scale observational diabetes data}

In order to check if the $ClinTrajan$ package can be applied to larger datasets, we extracted clinical trajectories using a publicly available dataset, representing 10 years (1999-2008) of clinical care at 130 US hospitals and integrated delivery networks. The dataset contains 101766 records satisfying the following conditions: (1) it is an inpatient encounter (a hospital admission), (2) it is a diabetic encounter, that is, one during which any kind of diabetes was entered to the system as a diagnosis, (3) the length of stay was at least 1 day and at most 14 days, (4) laboratory tests were performed during the encounter, (5) medications were administered during the encounter. The data contains such attributes as patient race, gender, age, admission type, time in hospital, medical specialty of admitting physician, number of lab test performed, HbA1c test result, diagnosis, number of medication, diabetic medications, number of outpatient, inpatient, and emergency visits in the year before the hospitalization. In the supervised setting, the aim of the analysis of this dataset is usually to predict the readmission event ('readmitted' variable) within 30 days after disposal from the hospital. In our analysis, we considered the readmitted variable as a part of the data space, in order to perform unsupervised analysis of the dataset with the aim to extract clinical trajectories, some of them leading to the increased readmission likelihood.

The exact protocol for encoding the diabetes dataset is provided at the $ClinTrajan$ github \url{https://github.com/auranic/ClinTrajan/}. Importantly, we encoded several categorical variables as ordinal. In particular, the readmitted variable was encoded in three levels with 0 value corresponding to 'No' (absence of recorded readmission), 1 - to '>30 days' and 2 - to '<30 days'. The `A1Cresult' feature (related to the HbA1c test) was encoded in two variables. The first one was binary indicating absence ('None' value) or presence of the measurement event. The second was the actual level of HbA1c: missing values corresponding to 'None', and three level encoding for the measured values, 0 - for 'Norm', 1 - for '>7' and 2 for '>8'. Since the A1Cresult field was not 'None' only in 17\% of patient records, this created a column containing 83\% of missing values, which were further imputed from the rest of the data. This was the only variable containing missing values. Age field was encoded as a 10-level ordinal variable accordingly to 10 age intervals provided in the initial data table.

For encoding the 23 categorical fields of the dataset describing the administered medications and change in their dosage, we used the following schema. First, we kept only four most frequently (in >10\% of cases) medications: insulin (53\% cases), metformin (20\% cases), glipizide (12\% cases) and glyburide (10.4\% cases). Second, each medication field was encoded into two variables: one binary indicating the absence ('No' value) or presence ('Steady' or 'Down' or 'Up') of the treatment prescription, and one three-level with 0 corresponding to either absence or no change in the treatment dose ('No' or 'Steady'), -1 corresponding to the decreased dose ('Down') and +1 to the increased dose ('Up'). 

We did not include some of the categorical variables such as admission type or diagnosis in the definition of the data space, since they contained hundreds of different values, and we used them rather as annotations to be visualized on top of the constructed tree. 2.3 \% of records corresponding to the elapsed states (hence, without possibility of readmission) have been excluded from the analysis, similarly to some previous analyses \cite{Long2018}.

The resulting encoded dataset contained 22 variables (8 numerical, 7 ordinal and 7 binary). We performed the data pre-processing similar to the one done in the infarction datasets, with imputing the missing values in the 'A1Cresult\_value' column and with further application of optimal scaling to ordinal values (see the corresponding Jupyter notebook at \url{https://github.com/auranic/ClinTrajan/}). The dimensionality of the dataset was reduced to 6 as it was the consensus value resulted from application of several methods of intrinsic dimension estimation (see Supplementary Figure 1), excluding outlying measurements.

\begin{figure*}[bt!] 
\centering
\includegraphics[width=\linewidth]{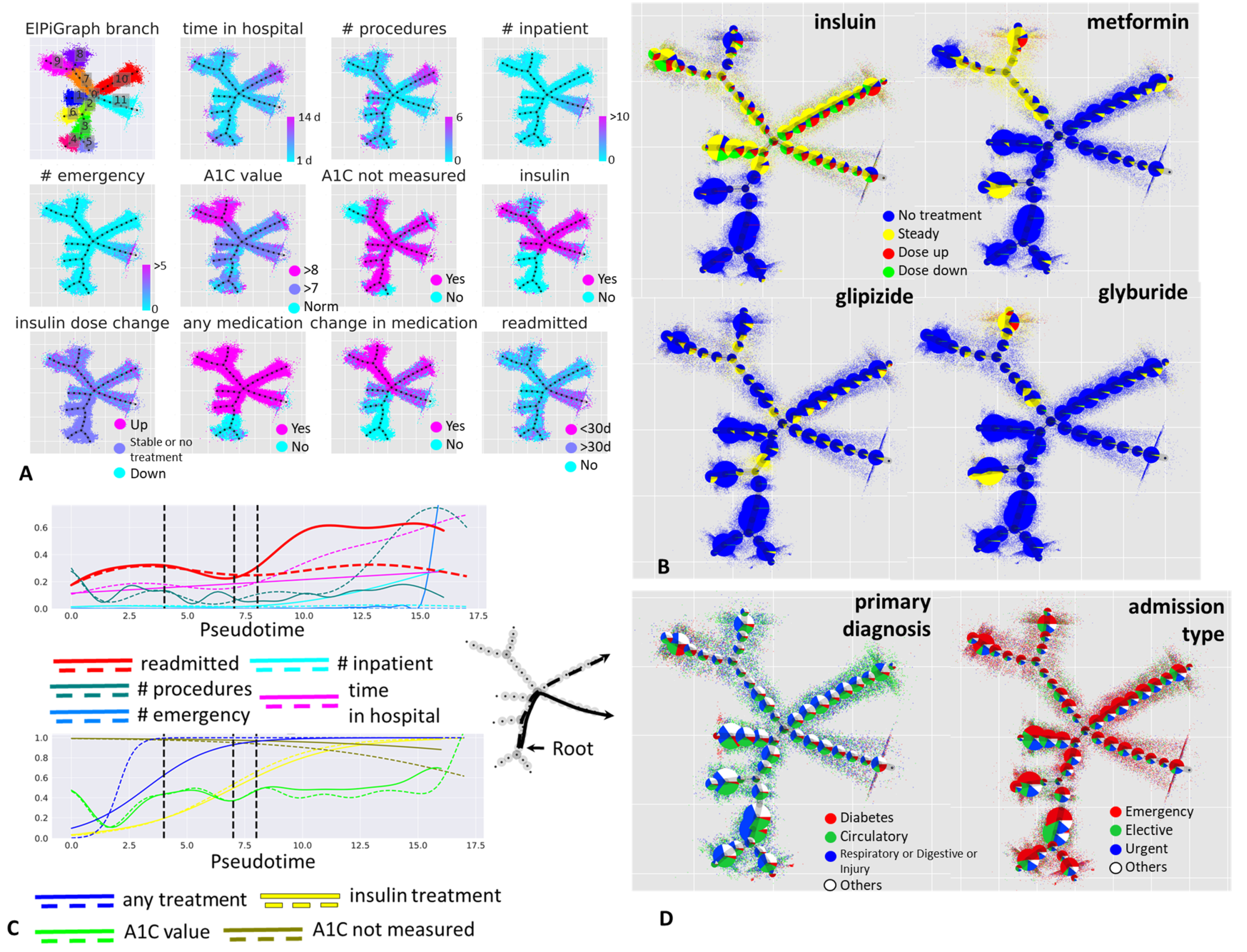}
\caption{Analysis of clinical trajectories in large-scale diabetes dataset. A) visualization of various clinical variables on top of the metro map layout of the principal tree. Partitioning the data accordingly to the principal tree segments is also shown in the top left corner. B) Visualization of four categorical variables related to the administered drug treatments and their dose changes. Here, the data points are shown in semi-transparent background, while on top of each graph node the relative proportions of the associated (the closest) data points are shown as a pie-chart. The size of the pie-charts are proportional to the number of points associated to each node. C) Pseudotemporal dynamics of clinical variables correlated to readmission- and long stay-associated clinical trajectories (shown as solid and dashed lines on the left). D) Visualization of the data table fields not participating in the construction of the principal tree, ``primary diagnosis" on the left and ``admission type" on the right.}\label{fig:DiabetesTrajBifurcations}
\end{figure*}

The principal tree algorithm was applied with the same parameters as in the previous section. The construction of the principal tree with 50 nodes for the 6-dimensional dataset with 99343 data points took approximately 400 seconds on an ordinary laptop. The principal tree explained 64\% of the total variance in contrast to 47\% of the first two principal components, with 4 PCs needed to explain the same percentage of variance as the principal tree. The tree contained 8 branching points (see Figure~\ref{fig:DiabetesTrajBifurcations},A) with one forth-order star. 
The principal tree-based data layout was used to visualize the values of data space variables (Figure~\ref{fig:DiabetesTrajBifurcations},A), and some other variables from the annotation data (Figure~\ref{fig:DiabetesTrajBifurcations},B,D), some of which did not participate in determining the structure of the principal tree. 

As a root node in this case, we selected the middle node of one of the internal segments of the principal tree (segment \#3 in Figure~\ref{fig:DiabetesTrajBifurcations},A), which was characterized by the shortest times spent in the hospital, smallest number of all procedures, no history of inpatient stays or emergency calls in the preceding year, normal predicted (not measured) value of HbA1C, absence of any medication. Therefore, this area of the principal tree was considered as corresponding to quasi-normal state in terms of diabetes treatment. 

Starting from this root node, the structure of the principal tree allowed us to define 8 distinct clinical trajectories. We focused on two of them, depicted in Figure~\ref{fig:InfarctusTrajBifurcations},C as solid and dashed lines, together with pseudo-time dependence of several selected clinical variables. One of these two trajectories was the only one associated to the high readmission incidence, increasing with pseudo-time. It did not correspond, however, to the longest stays in hospital, which was the feature of the second considered clinical trajectory. Therefore, we will designate these clinical trajectories as 'readmission-associated' and 'long stay-associated'. Not surprisingly, the readmission-associated trajectory was characterized by increasing number of inpatient and outpatient stays as well as the increasing number of emergency visits in the preceding year. This association must be interpreted by clinicists in order to attribute it either to objective clinical patient state requiring frequent return to the hospital or a psychological pattern of behaviour. In favour of the objective cause, one can notice that the readmission-associated trajectory contains different spectrum of primary diagnoses compared to the long stay-associated trajectory where the primary diagnoses related to circulatory system are dominating (Figure~\ref{fig:InfarctusTrajBifurcations},D,left). We can also notice the elective hospitalizations were increasingly more frequent for the long stay-associated trajectory, while the pseudo-time of the readmission-associated trajectory correlates with increasing probability of admission by emergency (Figure~\ref{fig:InfarctusTrajBifurcations},D, right).

Readmission-associated trajectory in this analysis can be considered as undesirable clinical scenario, the main source of burden on the medical system with respect to the diabetes disease. By the trajectory-based analysis we confirmed previous conclusions from \cite{Strack2014} that the readmission-associated trajectory was connected with almost complete absence of the measured HbA1C (Figure~\ref{fig:InfarctusTrajBifurcations},C), unlike long stay-related trajectory where up to 40\% of patients passed through HbA1C testing at the final pseudotime values. The predicted value of HbA1C along readmission-related trajectory was '>7' (moderate elevation). Both readmission- and long stay-associated trajectories were characterized by administered treatment by insulin, with slightly more metformin indications along the long stay-associated trajectory. Importantly, the long stay-associated clinical trajectory is connected to the earlier, in terms of pseudotime, ``any treatment" variable dynamics (Figure~\ref{fig:InfarctusTrajBifurcations},C,bottom panel).

\subsubsection{Trajectory-based analysis of the relation between early readmission rate and the measured glycated hemoglobin HbA1c}

In the original publication of the diabetis dataset, several observations have been reported \cite{Strack2014}. First, it was observed that the dependency of early readmission (in less than 30 days) frequency estimate on the fact of the measurement of HbA1c is conditional on the type of the primary diagnosis (with three major ones being diabetis, circulatory and respiratory diseases). Second, for the patients with diabetis as primary diagnosis, it was shown that not measuring the level of HbA1c is connected to the increased risk of early readmission. Interestingly, from the Figure~1 of \cite{Strack2014}, one can conclude that, in patients with diabetis as primary diagnosis, high levels of measured HbA1c are connected to decreased readmission risk compared to the normal level of measured HbA1c. This quite paradoxical observation was done in simple calculations of the early readmission frequency as well as in rate calculations adjusted for several clinical covariates. 

\begin{figure*}[bt!] 
\centering
\includegraphics[width=\linewidth]{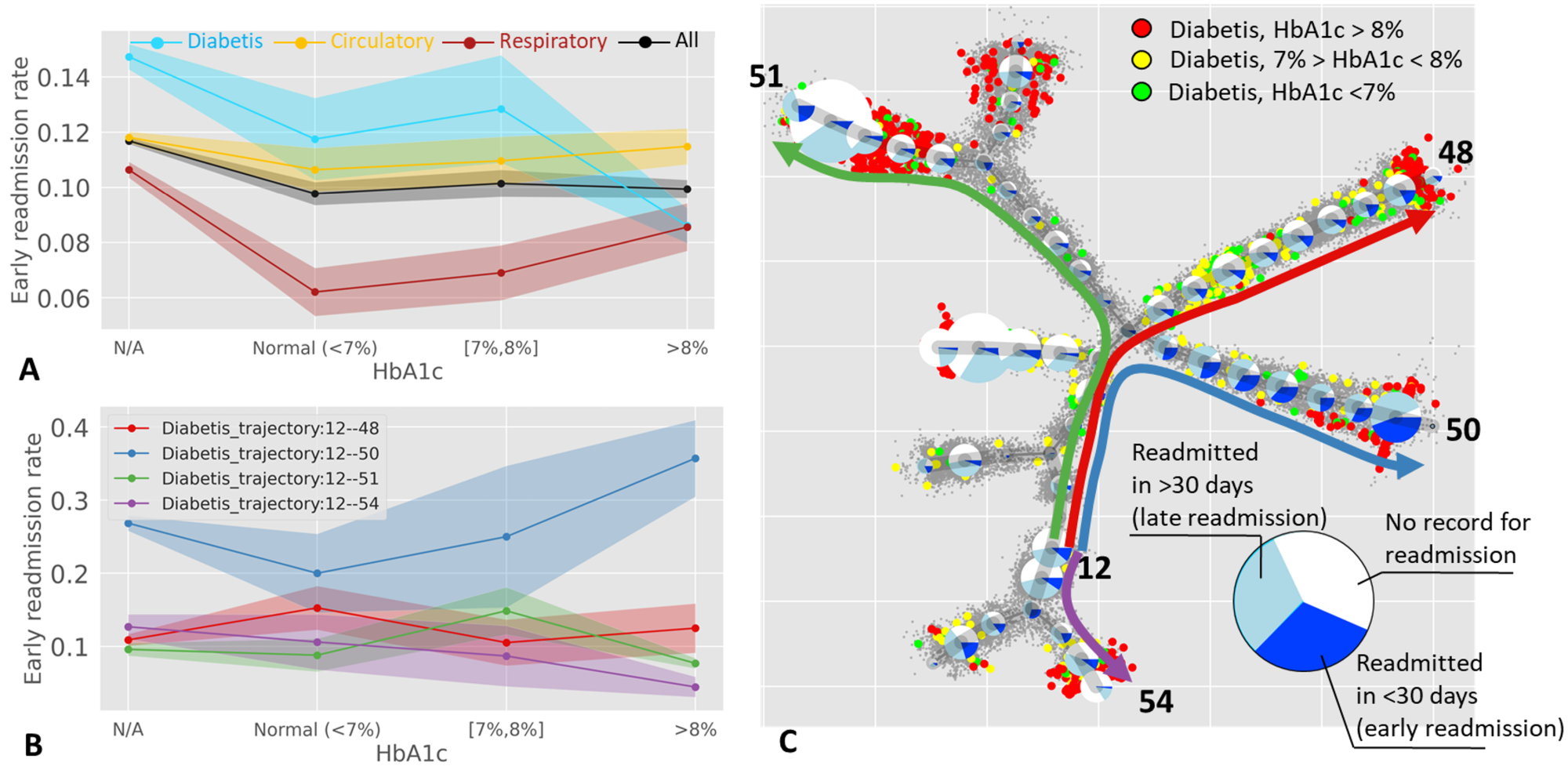}
\caption{Relation between early readmission and the measured glycated hemoglobin HbA1c. A) Global frequencies of early readmission (in less than 30 days) as a function of HbA1c measure. B) Frequencies of early readmission along four representative trajectories (shown in C panel), computed only for the patients with diabetis as primary diagnosis. Standard deviation interval is shown by shaded area for each curve in panels A and B. C)  Readmission frequencies shown along the clinical trajectories as pie-chart diagrams. The size of the pie-charts reflects the number of patients with diabetis as primary diagnosis associated with each node of the principal tree. Bigger color data points show patients with diabetis as primary diagnosis, with known measured level of HbA1c (normal in green, medium in yellow and high in red). Note that heavier cases of diabetis confirmed by HbA1c tend to be at the end of the clinical trajectories. The colors of the trajectories  in C) and the plots in B) are matched.}\label{fig:DiabetesReadmission}
\end{figure*}

In order to illustrate the advantage of the trajectory-based patient stratification, we recomputed the simple not adjusted estimations of the early readmission frequency as a function of measured HbA1c in sets of patients with different primary diagnosis (Figure~\ref{fig:DiabetesReadmission},A). This reproduced the previously made conclusions from the original study \cite{Strack2014}. The frequency of readmission appeared to be higher in the patients with not measured HbA1c. Qualitatively similar to the previous publication, the readmission rate was significantly lower for the high values of HbA1c compared to its normal levels (8.6\% vs 11.8\%). Note that the analyzed dataset has changed since it original publication with more than 20000 new patients being added. In order to explain this seeming paradox, we hypothetized that it can be explained by the heterogeneity of relations between the levels of HbA1c and readmission, which can be captured in distinct clinical trajectories. 

We looked at the cases of primary diagnosis of diabetis separately for each 8 clinical trajectories previously identified via the principal tree method application (Figure~\ref{fig:DiabetesReadmission},B). Quite strikingly, the dependence of early readmission on the measured levels of  HbA1c is clearly different along different trajectories (Figure~\ref{fig:DiabetesReadmission},C). For example, the trajectory ending with node '50' (Figure~\ref{fig:DiabetesReadmission},B and C, denoted as 'Trajectory 12-50'), was associated with higher risks of readmission. Absence of HbA1c measurement is still associated with higher level of early readmission (26.8\% compared to 20\% for the cases with normal HbA1c level). However, along this trajectory the higher levels of measured HbA1c are associated with much higher levels of readmission (35.7\%). There exist another trajectory ('Trajectory 12-54'), where the dependence follows an opposite pattern (12.7\% of early readmissions for not measured HbA1c against 10.6\% for normal levels of HbA1c and 4.4\% for high levels of HbA1c). 

Therefore, we can tentatively suggests that different trajectories in the diabetis data stratify the patients into clinically distinct scenarios, requiring different statistical models for anticipating the readmission rates. As a consequence, measuring HbA1c might have more clinical value in terms of estimating risks of early readmission along some trajectories and less along the others. For example, the patients with diabetis as primary diagnosis along the trajectory 'Trajectory 12-50' are characterized by frequent readmissions, with heavier cases of diabetis leading to very frequent reamissions. Two other trajectories exemplified in Figure~\ref{fig:DiabetesReadmission},B-C, show much less effect of the measurement of HbA1c on readmission rates, and one trajectory show an opposite trend, with heavier cases being less frequently readmitted. As a consequence, we can suggest that measuring the level of HbA1c is critical for determining the risk of early readmision for 'Trajectory 12-50' and 'Trajectory 12-54', but appears to be less important for  'Trajectory 12-48' and 'Trajectory 12-51'. 

Interpretation of distinct clinical scenarios must be performed by experts in the field of diabetis treatment. We can only hypothesize that the seeming drop in the globally assessed readmission rates in patients with high measured HbA1c might be connected to existence of a large subset of 'stabilized' patients, with established supportive treatment. However, the stratification of patients into different clinical trajectories demonstrates that this is not a universal effect, and that one can distinguish other patient clusters characterized by heavy forms of diabetis characterized by relatively high rates of early readmission. 

\section{Discussion}

In this study we considered two rich and large publicly available observational clinical datasets from the two most challenging areas of public health: cardiology and diabetes. Both datasets contain syncronic (related to the moment of staying in the hospital) observations over a relatively large population of patients. Therefore, the traditional unsupervised machine learning approach for treating these data in order to classify clinical states is supposed to be some kind of clustering or manifold learning. We demonstrate that there exists an alternative approach which allowed us to represent these data as pseudo-diachronic, i.e., reflect to some extent the temporal aspects. This opens a possibility to classify not only the states of particular patients but their hypothetical clinical trajectories arriving from the past and projected into the future. This in turn gives a possibility to reason in terms of dynamical disease phenotyping, e.g., classifying clinical states in terms of the disease dynamics type. 

Identification of clinical trajectories is made possible due to the use of the branching pseudotime approach, consisting in modeling the geometry of the dataset as ``bouquet" of diverging trajectories, starting from one or several hypothetical quasi-normal, e.g., characterized by the least severe condition, disease states. The progression along a particular clinical trajectory can be quantified in terms of pseudotime, reflecting the abstract accumulated amount of changes in the observed clinical traits. The main requirement for the possibility of such reconstruction is the existence of sufficient number of observations (thousands) such that the individual variations in the clinical states would reveal the major non-linear routes along which they progress in real physical time.

Trajectory analysis from the snapshot data is a widely used approach in modern molecular single cell studies, where the genome-wide measurements of individual cell states are inevitably destructive. Collecting the information about a large number of cell states allows reconstructing the underlying hidden cellular dynamics without following each individual cell in physical time \cite{Chen2019}. Dynamic phenotyping of cell states is a rapidly emerging concept in this scientific field \cite{Ruderman2017}. Elastic principal graphs (ElPiGraph) is an established general machine learning method which is widely used for the purpose of reconstructing cellular trajectories from the single cell data, in the form of principal trees or other more or less complex graph topologies \cite{Albergante2020}. Here we suggest to apply ElPiGraph to quantification of clinical trajectories in large clinical datasets, which requires adapting ElPiGraph to the datasets characterized by mixed data types and presence of non-randomly distributed missing values.

This effort resulted in $ClinTrajan$ Python package which can be readily applied in the analysis of clinical datasets containing even millions of observations. In the real life diabetes dataset considered here and containing more than one hundred thousands of observations, the analysis by $ClinTrajan$ takes few minutes on an ordinary laptop.

Use of ElPiGraph is the most relevant in the case when the hypothetical probability density function underlying the multi-dimensional data is characterized by certain archetypal features. Recall that classical phenotyping is, in its essence, cluster analysis of data. Application of standard clustering methods assumes existence of lumps and peaks in the density function: therefore, clustering looks for a set of principal points \cite{Xu2008,Tarpey1996}. 

Dynamical phenotyping has a different basic assumption that the point density is characterized by the existence of continuous one-dimensional ``ridges" which can diverge from or converge to each other in the data space. They can also connect local density peaks. In this case, the appropriate data approximation methods (such as ElPiGraph) look for principal curves and, more generally, branching principal trajectories, along which the data points are condensed\cite{Hastie1989, Albergante2020}. The relevance of such a data model for dynamical phenotyping follows from the nature of a complex dynamical process, underlying disease progression, which develops in physical time and is sampled in the space of clinical characteristics.


Similar to the cellular trajectories, the reconstructed clinical trajectories do not possess any natural orientation: therefore, orienting them requires expert-based decisions for choosing one or several root nodes in the principal tree. Also, the hypothetical dynamics of patients along the clinical trajectory does not have to be assumed irreversible. Some additional insights about orientation and reversibility can be obtained from a mix of synchronic and diachronic data, where individual patients can be represented not by simple data points but by the more or less longitudinal observations represented by short trajectories. The best practices of using such data from the machine learning perspective remains to be established \cite{Nagin2010}.

It appears interesting to relate the inferred pseudotime with the physical time and use it to parametrize the obtained clinical trajectories. This appears for us an important challenge which can be approached in several ways.

One of them is related to the aging of patients. Indeed, the chronological age represents the most basic way to rank the patients in a sequence which can potentially correlate to the clinical state (hence, define a clinical trajectory). However, the relation between 'biological' and 'chronological' age remains complex, especially in the pathological context \cite{Whitwell2020}. In our study we exploited the chronological patient age as any other clinical variable, and observed that indeed age correlates to some clinical trajectories but not to the others. Moreover, some clinical trajectories might be characterized by decreasing chronological age, which can be interpreted as aggravating clinical picture specific to younger patients. We can imaging other ways of using the age variable: for example, for learning the structure of the principal tree in a semi-supervised fashion. How to use the chronological age in the most informative way when analyzing both longitudinal and synchronic data remains an open question \cite{Whitwell2020}.

Second approach to introduce physical time in the picture is using partially diachronic data as an additional annotation of a clinical dataset (the case of complete diachronic clinical data, representing longitudinal observations, is usually treated using a different and established set of approaches). One source of information which can be relatively easily obtained is identifying pairs of data points corresponding to two subsequent states of the same patient, and recording the time lapse between two states. For example, part of the patients in a clinical dataset can be returning to hospital, with a previous record included in the dataset, so this information must be available. If the number of such pairs is sufficiently large then one can try to learn a monotonic function $F_k$ of pseudotime along each trajectory $k$, predicting the actual temporal label for each observation. Note that the connection between pseudotime and physical time can be different along different clinical trajectories. Moreover, the paired patient observation data can be used in the process of principal tree learning, by minimizing the number of paired patient observations belonging to distinct clinical trajectories.

Another limitation of the suggested approach is that the clinical trajectories are assumed to be diverging from some initial root state or states. In reality, convergence of clinical trajectories seems to be feasible (as in the case of the cellular trajectories). In this case, the model of the principal tree has to be generalized to some more general graph topologies (e.g., existence of few loops). In case of ElPiGraph method, such modifications are easy to introduce technically: however, introducing graph structures more complex than trees requires careful consideration in order to avoid creating data approximators whose complexity will be comparable to the complexity of the data themselves \cite{Zinovyev2013}.

\section{Potential implications}

Quantification of clinical trajectories represents the first step in using the concept of dynamic clinical phenotyping for diagnostics and prognosis. Predicting the probabilities of future clinical states for a particular patient together with their uncertainties, using the knowledge of clinical trajectories, can be a natural next step for future studies. These approaches can consider clinical trajectories as a coarse-grained reconstruction of the state transition graph for a dynamic system, described by, for example, continuous Markov chain equations. Some methodological ideas can be borrowed from the recent omics data studies \cite{Setty2019}.


Recapitulating the multi-dimensional geometry of a clinical dataset in terms of clinical trajectories might open possibilities for more efficient applications of other methods, more oriented towards supervised machine learning. For example, it can be potentially used for learning the optimal treatment policy, based on application of reinforcement learning as in \cite{Saria2018}.

The existing large clinical datasets are frequently collected as a result of multi-site studies. In case of strong artefacts and biases caused by application of significantly different practices for data collection or other factors, specific methods of correction should be applied, integrated into the data analysis or even in the study design\cite{Chen2012Design}. However,  dimensionality reduction methods based on averaging can in principle partially compensate for data heterogeneity if it can be modeled as a mixture of independent site effects that remain relatively small compared to the ranges of variable variations along the clinical trajectories. ElPiGraph in this respect has advantages over much more rigid Principal Component Analysis, being a nonlinear generalization of it for the case of datasets with complex geometries\cite{Albergante2020}. However, this aspect of ElPiGraph requires a specific further investigation.

Overall, we believe that introducing trajectory-based methodology in the analysis of synchronic datasets might change the angle of view on their use for developing prognostic and diagnostic expert systems.

\section{Methods}

\subsection{Implementation of the methodology}

The $ClinTrajan$ methodology is implemented in Python, packaged and openly available at \url{https://github.com/auranic/ClinTrajan/} together with Jupyter notebooks providing the exact protocols of applying the $ClinTrajan$ package to several case studies. The detailed description of $ClinTrajan$ functionality is provided from its web-site. 


\subsection{Quantification of mixed type datasets}

Quantification of mixed type datasets, i.e. assigning a numerical value for nominal variables is a vast field where many solutions have been suggested \cite{Young1981}. In the $ClinTrajan$ package we used several popular ideas adapted to the aim of finding non-linear trajectories in the data. 

Firstly, for all non-binary categorical variables we suggest applying the ``dummy'' encoding (or ``one-hot'' encoding), e.g. introducing new columns, containing binary values one per each  category (one of the categories might be dropped as redundant). Alternatively, if there is a sufficient number of numerical variables, Categorical Principal Component Analysis (CatPCA) can be applied \cite{Linting2012, Casacci2015}. 

Secondly, for ordinal variables (including binary ones as particular case) we suggest to use either univariate or multivariate quantification. For univariate quantification we assume that the ordinal values are obtained by binning a ‘latent’ numeric continuous variable possessing the standard normal distribution (zero mean and unit variance), following the approach described in \cite{Fehrman2019}. Let us consider an ordinal variable $V$ which takes ordered values $v_1<v_2\dots<v_m$, and each $v_i$ value has $n_i$ counts in the dataset. We quantify $V$ by the values

\begin{equation}
x_i = \Phi^{-1}\left( \sum_{j=1}^{i-1}p_j+\frac{p_i}{2}\right),
\end{equation}

\noindent where $p_i=\frac{n_i}{N}$, $N$ is the total number of data points, and $\Phi(x) = \frac{1}{\sqrt{2\pi}}\int_{-\infty}^xe^{-\frac{x^2}{2}}dx$.

If there exist many ordinal variables in the dataset, one can use multivariate methods for joint quantification of them. One of the most popular approaches is a particular variant of {\it optimal scaling}, aiming at maximizing the sum of squared pairwise correlations between all variables, including numerical and ordinal ones \cite{Young1981}. $ClinTrajan$ package includes its own implementation of this variant of optimal scaling which can be used to quantify ordinal variables in clinical datasets. 

The advantage of multivariate ordinal variable quantification with respect to the univariate one in that it can decrease the intrinsic dimensionality of the resulting data point cloud which can be beneficial for further applying of manifold learning methods, including the method of elastic principal graphs (ElPiGraph). The disadvantage of multivariate quantification of ordinal variables consists in necessity to have sufficiently large portion of data table rows without missing values. If this part is small then it might be not possible to quantify certain ordinal levels since they won't be represented in this complete part of the dataset, while this might still be possible with univariate quantification. 

Thus, imputing missing values requires quantification of ordinal variables, and multivarite quantification of them requires imputation of missing values. Therefore, in practice we apply a hybrid approach consisting in application of univariate quantification with further imputation of missing values and further application of multivariate quantification using the optimal scaling approach.

Lastly, we suggest to transform all continuous numerical variables to the standard $z$-scores (i.e., centering and scaling), in order to make them comparable.

\subsection{Imputing missing values in mixed type datasets}

The real-life clinical datasets are almost always only partially complete and contain missing values. Typically, these values are not distributed uniformly across rows and columns of the data matrix but rather form some non-random patterns, which can be even constructively used for the tasks of clinical data analysis \cite{Mirkes2016}. A typical pattern is existence of a column (or a row) containing abnormally large number of missing values. One can define two parameters $\delta_{row}$ and $\delta_{column}$ as maximally tolerable fraction of missing values in any row or column of the data matrix. The problem of finding the maximum size sub-matrix satisfying these constraints is not completely trivial but can be approximated by some simple iterative approaches. In practice, the trivial suboptimal solution consists in eliminating columns having the fraction of missing values larger than $\delta_{column}$, and then eliminating the rows having the fraction of missing values larger than $\delta_{row}$.

After constraining the maximum fraction of missing values in the data matrix, one can apply one of the available missing value imputation algorithms (imputers), which can be also classified into univariate and multivariate. For our purposes we advocate the use of multivariate imputers that allow us to avoid having strong data outliers destroying the manifold structure of the dataset. The standard scikit-learn collection provides two types of imputers: nearest neighbors imputation and iterative multivariate one, which can be in principle used for this purpose. In $ClinTrajan$ package We add two alternative imputers based on application of Singular Value Decomposition (SVD) of order $k$. The first one which we will designate as ``SVDComplete" is applicable if the number of rows in the data matrix having no missing values is sufficiently large (e.g., not much smaller than 50\%). Then the standard SVD or order $k$ is computed on the sub-matrix having only complete rows, and each data vector containing missing values is projected into the closest point of the hyperplane spanned by the first $k$ principal components. The imputed value is then read out from the projected vector. For ordinal and binary variables, the imputed value can be additionally rounded to the closest discrete numerical value, in order to avoid ``fuzzy values'' which do not correspond to any initial nominal value. The mutually exclusivity of binary variables encoding the categorical fields can be also taken into the account. The second SVD-based imputer is called ``SVDFull'' and it is based on computing SVD of order $k$ for the full matrix with missing values; for example, using the method suggested in \cite{Dergachev2001, Gorban2008Principal}. After computing the principal vectors, the imputation is performed as in ``SVDComplete" imputer. Choice of $k$ can be made either through applying cross-validation, or using a simple heuristics consisting in setting $k$ to the value of the intrinsic dimensionality (ID) of the data. ID can be estimated through the application of full-order SVD and analyzing the scree plot, or through a number of more sophisticated approaches \cite{Albergante2019}.

\subsection{Method of Elastic Principal Graphs (ElPiGraph)}\label{elpigraph_method}

\subsubsection{Computing the elastic principal graph}

Elastic principal graphs are structured data approximators \cite{Gorban2007Topological, Gorban2008book, Gorban2010Principal, Gorban2008Principal}, consisting of nodes connected by edges. The graph nodes are embedded into the space of the data, minimizing the mean squared distance (MSD) to the data points, similarly to the k-means clustering algorithm. However, unlike unstructured k-means, the edges connecting the nodes are used to define the elastic energy term. This term is used to create penalties for edge stretching and bending of segments. To find the optimal graph structure, ElPiGraph uses topological grammar (or, graph grammar) approach and gradient descent-based optimization of the graph topology, in the set of graph topologies which can be generated by a limited number of graph grammar operations. 

Elastic principal graph is an undirected graph with a set of nodes $V=\{V_i\}$ and a set of edges $E=\{E^i\}$. The set of nodes $V$ is embedded in the multidimensional space. In order to denote the position of the node in the data space, we will use the notation $\phi(V_j)$, where $\phi(V_j)$ is a map $\phi:V\rightarrow R^m$. The optimization algorithm search for such $\phi()$ that the sum of the data approximation term and the graph elastic energy is minimized. The optimization functional is defined as:

\begin{equation}
U^{\phi}(X,G) = MSD^{\phi}(X,V)+U^{\phi}_E(G)+U^{\phi}_R(G),
\end{equation}

\noindent where

\begin{equation}
MSD^{\phi}(X,V) = \frac{1}{|X|}\sum_{i=1}^{|X|}\min(||X_i-\phi(V_{P(i)})||^2,R_0^2),
\end{equation}

\begin{equation}
U^{\phi}_E(G)=\sum_{E^{i}}\lambda_{penalized}(E^{i})\left(\phi(E^{i}(0))-\phi(E^{i}(1))\right)^2,
\end{equation}

\begin{equation}
U^{\phi}_R(G)=\mu \sum_{S^{j}} \left(\phi(S^{j}(0))-\frac{1}{deg(S^{j}(0))}\sum_{i=1}^{deg(S^{j}(0))}\phi(S^{j}(i)) \right)^2,
\end{equation}

\begin{equation}
\lambda_{penalized}(E^{i}) = \lambda+\alpha\left( \max(2,deg(E^{i}(0)),deg(E^{i}(1)))
-2\right),
\end{equation}

\noindent where $|V|$ is the number of elements in set $V$, $X = \{X_i\},i = 1,\dots,|X|$ is the set of data points, $E^i(0)$ and $E^i(1)$ denote the two nodes of a graph edge $E^i$, star $S^j$ is a subgraph with central node $S^j(0)$ and several (more than 1) connected nodes (leaves), $S^j(0),\dots,S^j(k)$ denote the nodes of a star $S^j$ in the graph (where $S^j(0)$ is the central node, to which all other nodes are connected), $deg(V_i)$ is a function returning the order $k$ of the star with the central node $V_i$, and $P(i) = \text{argmin}_{j=1,\dots,|V|}\|X_i-\phi(V_j)\|^2$ is a data point partitioning function associating each data point $X_i$ to the closest graph node $V_{P(i)}$. $R_0$, $\lambda$, $\mu$, and $\alpha$ are parameters having the following meaning: $R_0$ is the trimming radius such that points further than $R_0$ from any node do not contribute to the optimization of the graph, $\lambda$ is the edge stretching elasticity modulo regularizing the total length of the graph edges and making their distribution close to equidistant in the multidimensional space, $\mu$ is the star bending elasticity modulo controlling the deviation of the graph stars from harmonic configurations (for any star $S^j$, if the embedding of its central node coincides with the mean of its leaves embedding, the configuration is considered harmonic). $\alpha$ is a coefficient of penalty for the topological complexity (existence of higher-order branchings) of the resulting graph.

Given a set of data points and a principal graph with nodes embedded into the original data space, a local minimum of $U\phi(X,G)$ can found by applying a splitting-type algorithm. Briefly, at each iteration given the initial guess of $\phi$, the partitioning $P(i)$ is computed, and then, given the $P(i)$, $U\phi(X,G)$ is minimized by finding new node positions in the data space. A remarkable feature of ElPigraph is that the $U\phi(X,G)$ minimization problem is quadratic with respect to node coordinates and is reduced to solving a system of linear equations. Importantly, the convergence of this algorithm is proven \cite{Gorban2008Principal, Gorban2017Robust}.

Topological grammar rules define a set of possible transformations of the current graph topology. The graph configuration of this set possessing the minimal energy $U\phi(X,G)$ after fitting the candidate graph structures to the data is chosen as the locally best with a given number of nodes. Topological grammars are iteratively applied to the selected graph until given conditions are met (e.g., a fixed number of grammar application, or a given number of nodes is reached, or the required approximation accuracy $MSD^{\phi}(X,V)$ is achieved). The graph learning process is reminiscent to a gradient descent-based optimization in the space of all possible graph structures achievable by applying a set of topological grammar rules (e.g., in the set of all possible trees).

One of the simplest graph grammars consists of two operations 'add a node to node' and 'bisect an edge', which generates a discrete space of tree-like graphs \cite{Albergante2020}. The resulting elastic principal graphs are called \textit{elastic principal trees} in this case. In $ClinTranjan$ package we currently use only principal trees for quantifying trajectories and pseudotime, even though using more general graph topologies is possible. The advantages of limiting of the graph topology to trees are in that it is easy to layout the structure of the graph on a 2D plane and that any trajectory connecting two nodes of the graph is unique.

The resulting explicit tree structure can be studied independently on the data. Also, an artbitrary vector $x$ -- not necessary belonging to the dataset $X$ -- can be projected onto the tree and receive a position in its intrinsic geodesic coordinates. The projection is achieved by finding the closest point on the principal graph as a piecewise linear manifold, composed of nodes and edges as linear segments connecting nodes. Therefore, the projection can end up in a node or on an edge. In further we define a projection function $\{p,\epsilon \} = Proj(x,G)$, returning a couple containing the index of the edge which is the closest one to $x$ and the position of the projection from the beginning of the edge $E^p(0)$ as a fraction of the edge length $\epsilon \in [0,1]$. Therefore, if $\epsilon=0$ then $x$ is projected into  $E^p(0)$ and if $\epsilon=1$ then the projection is in $E^p(1)$. If $\epsilon \in (0,1)$ then the projection is on a linear segment, connecting $E^p(0)$ and $E^p(1)$.

A detailed description of ElPiGraph and related elastic principal graph approaches is available elsewhere \cite{Albergante2020}. The ElPiGraph package implemented in Python is available from \url{https://github.com/sysbio-curie/ElPiGraph.P}. Implementations of ElPiGraph in R, Matlab, Java and Scala are also available from \url{https://sysbio-curie.github.io/elpigraph/index.html}.
When analyzing the clinical datasets, the principal tree inference with ElPiGraph was performed using the following parameters: $R_0 = \infty$, $\alpha = 0.01$, $\mu = 0.1$, $\lambda = 0.05$. After the initial principal tree was constructed, it was pruned and the terminal segments were extended. The pruning consists in eliminating the final terminal segments of the tree containing only one single edge. Extending the terminal segments consists in extrapolating the segment in order to have most of the data points projected on the edges of the terminal segment and not at its terminal node. Both functions are standard principal tree post-processing choices, implemented in ElPiGraph package.

\subsubsection{Partitioning (clustering) the data according to the principal graph segments}

Embedding a graph to the data space allows us to partition (cluster) the dataset in several natural ways: for example by assigning each datapoint to the closest node or the closest edge. However, these ways do not fully suit our purposes, since they do not reflect the intuition of ``trajectory''. So it is natural first to decompose  the graph itself into linear fragments without branching (we will call them non-branching graph segments or simply segments) and afterwards to cluster the dataset according to the closest segment. This is the idea of the data partitioning used in the paper, and it is described with more details below. 

By the branching node in a graph we denote any node with connectivity degree larger than 2 ($deg(V_i)>2$), and by the leaf or terminal node of the graph we denote a node with degree less than 2 ($deg(V_i)<2$).

\begin{figure}[H] 
\centering
\includegraphics[width=8cm]{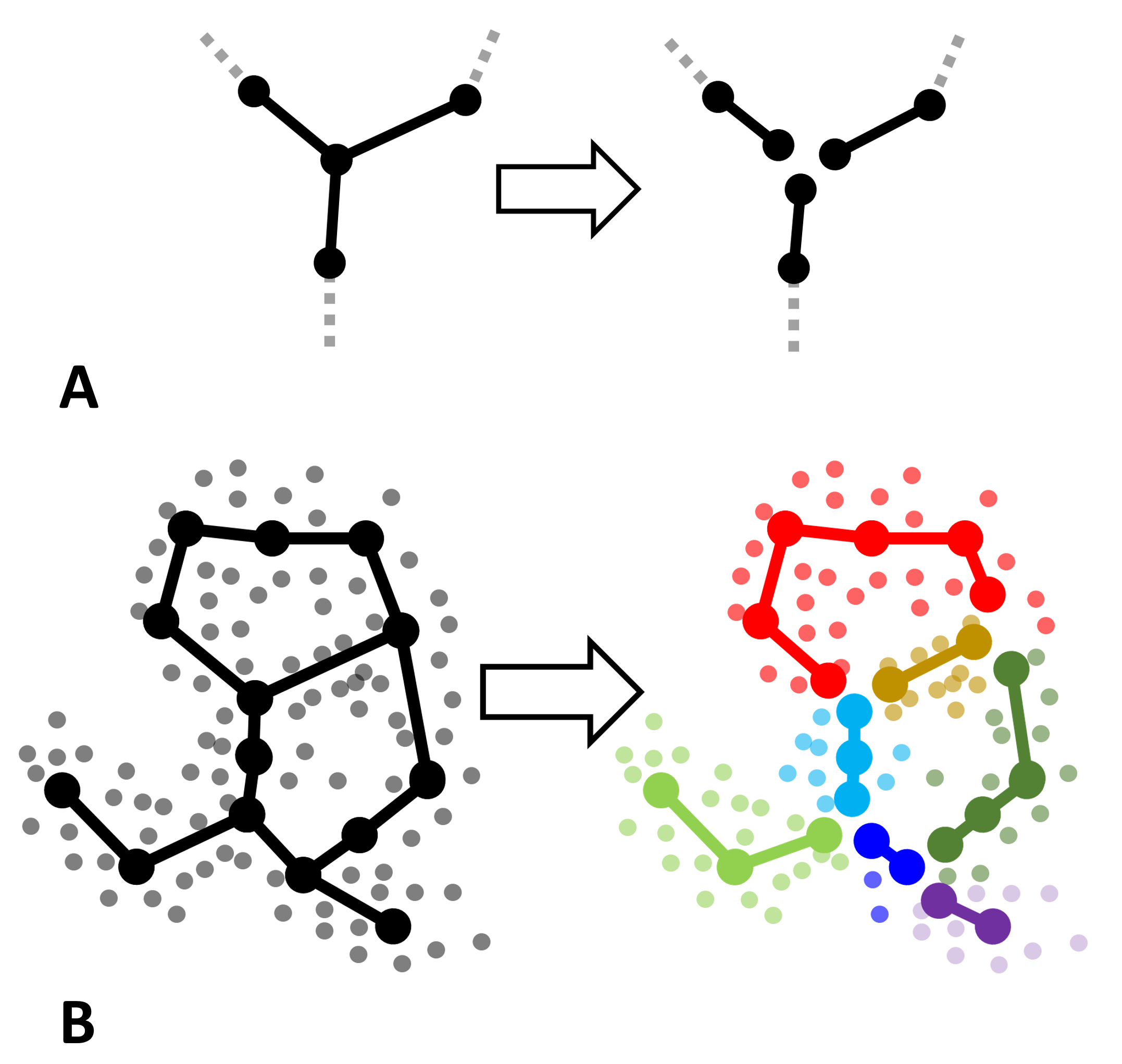}
\caption{Decomposing a graph into non-branching segments and partitioning the data accordingly to the principal graph segments. A) The principal operation for segmenting the graph: each branching point of the graph (having degree more than two) is multiplied and attached to the end of every edge composing the branching point such that the edges of the graph star are unglued and become disconnected. B) Toy example of a principal graph approximating a cloud of data points (shown in grey), and using its decomposition into non-branching segments for partitioning the data which can be considered a kind of clustering (see text for details).}\label{fig:GraphSegments}
\end{figure}

Let us call linear segment (or, simply segment) of a graph such a path which connects one branching node to another branching or a leaf node and which does not contain any other branching nodes. Internal segments connect two branching nodes and the terminal segments connect a branching node to a leaf node (Figure~\ref{fig:GraphSegments},A). As one can see such definition reflects the intuition underlying the notion of the ``segment''; we only need to specify several exceptional cases. For a graph which is an isolated cycle (not containing branching or leaf nodes), the whole cycle should be considered as a ``segment''. The same is true if a graph contains several connected components which are cycles: then all of them are considered as separate ``segments''. The other exceptional case are nodes of degree zero (isolated nodes) - they also will be considered as separate ``segments''. 
These exceptional cases cannot happen for connected principal trees which are the main object of the present study, they were just mentioned for completeness. 

Any graph can be uniquely split into ``segments'', which is not difficult to prove, especially for trees. 

We coded in Python a version of the depth-first search algorithm to produce a split into segments for an arbitrary graph. Main difference to the classical depth-first search is a storage of visited edges (not only nodes) of the graph to correctly process possible cycles in the graph. 
The algorithm starts from any branching or leaf node, and walks along edges in depth, joining them to the ``current segment'' until it meets a branching or a leaf node. Here, the ``current segment'' is terminated. In case of a leaf node one returns from the recursion, the same for already visited branching node. In case of a new branching node (not visited before) one goes into deeper level of depth-first recursive process. 
After all edges of a graph are partitioned into segments,
one can partition (or, cluster) the dataset accordingly to the closest segment which can be done in two ways. Firstly, choose nearest edge to a given datapoint and associate the point to the segment to which that edge belongs to. However, a simpler approach is much more computationally efficient: calculate distances from datapoints to nodes and choose the segment which contains the nearest node. In case this node belongs to several segments (therefore, it is a branching node), we choose the segment which contains the second nearest node among all nodes which belong to the corresponding segments. If the number of nodes in the graph is large enough then both approaches will produce (almost) identical results (Figure~\ref{fig:GraphSegments},B).

\subsubsection{Dimensionality reduction and data visualization using principal graphs}

To visualize the principal graph, each datapoint is first associated with the closest ElPiGraph edge in full dimensional space, and the distance to the projection onto the edge is recorded.

We then embed the graph structure in 2D by computing a force-directed layout with the Kamada-Kawai algorithm \cite{kamada1989algorithm}. Each datapoint is placed orthogonally on a random side of its associated edge, at the distance proportional to the distance to the projection in the initial space.
The proportionality constant is called scattering parameter, which is adjusted by a user, or can be optimized in order to preserve, in the best way, the structure of the distances betweeen the datapoints in the initial data space.

Edge widths can also be used to visualize the values of a variable or any function of the variables defined in the nodes of the graph.

\subsubsection{Quantifying pseudo-time and extracting trajectories using principal trees}

After computing the principal tree, a root node $V_{root}$ has to be defined by the user, accordingly to the application-specific criteria. For example, it can correspond to the node of the graph closest to a set of data points enriched with those having the least of disease severity.

The pseudotime $Pt(x)$ of an arbitrary vector $x$ is defined as the total geodesic distance in the principal tree from $V_{root}$ to the projection $\{p,\epsilon\}=Proj(x,G)$ of $x$ on the graph. Algorithmically, we need to define which node of the edge $E^p$ is the closest to the $V_{root}$ and add the $\epsilon$ accordingly, i.e.

\begin{equation}\label{pseudo_time}
Pt(x) = \left\{
                \begin{array}{ll}
                  |V_{root}\rightarrow E^p(0)|+\epsilon,& \text{if } |V_{root}\rightarrow E^p(0)|<|V_{root}\rightarrow E^p(1)|\\
                  |V_{root}\rightarrow E^p(0)|-\epsilon,& \text{if } |V_{root}\rightarrow E^p(0)|>|V_{root}\rightarrow E^p(1)|
                \end{array}
              \right.
\end{equation}

\noindent where $|V_i\rightarrow V_j|$ signifies the number of edges (length) of the trajectory $V_i\rightarrow V_j$. 

\subsection{Associating class labels and data variables and principal tree segments}

Segment labeling of the data points induced by the structure of the principal graph represent a categorical label which can be associated to the dataset variables of various types. 

In order to test if there is an association of the tree segments to a categorical variable (including binary as a particular case), we used the standard independence ${\chi}^2$ test. If the test was significant then we identified those segments which have the most unexpected value of the variable $k$ by considering a simple deviation score:

\begin{equation}\label{deviation_score}
Deviation_{k}(\text{value } j, \text{segment} i) = \frac{E^i_{kj} - O^i_{kj}}{E^i_{kj}},
\end{equation}

\noindent where $O^i_{kj}$ is the observed number of data points associated to the segment $i$ having value $j$ of the variable $k$ and $E^i_{kj}$ is the expected number of occurrences of the value $j$ of the variable $k$, from the standard independence assumption. Positive values of this score correspond to the positive enrichment and negative values for negative enrichment.

In order to test statistical association between tree segments and numerical variable (including ordinal as particular case), we used the standard ANOVA test representing the independent tree segment variable through the standard one hot encoding into a set of binary variables. If the test was significant then we evaluated the significance of each of the segments by looking at the value and the p-values of the generalized linear model coefficients for each segment. Positive values of the coefficients correspond to positive enrichment, and negative for negative enrichment.

\subsection{Associating data variables and trajectories}

For computing the score of association between a data variable $k$ and a trajectory, we compute the $R^2$ score of the regression:

\begin{equation}
x^k = F(Pt(x)), \text{for } x \in  X_{V_{root} \rightarrow V_j},
\end{equation}

\noindent where $V_j$ is one of the leaf node in the tree and $Pt(x)$ is the pseudotime value of the data point $x$ computed from (\ref{pseudo_time}). For continuous variables, $F()$ can be linear or a non-linear regression (for example, the most popular Gaussian kernel regression). For binary variables, we fit $F()$ by computing the logistic regression. We consider a variable $k$ associated to the trajectory $X_{V_{root} \rightarrow V_j}$ if $R^2$ of the regression problem solution exceeds certain threshold.

\subsection{Pseudotime survival analysis}

The survival analysis shown in Figure~\ref{fig:InfarctusSurvivalAnalysis} was performed using Python package 'lifelines' \url{https://github.com/CamDavidsonPilon/lifelines}. In order to estimate the hazard rate, we used non-parametric Nelson–Aalen estimator of the cumulative hazard rate function implemented in the package. This estimator uses the formula \cite{Nelson1969}:

\begin{equation}
H(t) = \sum_{t_i\leq t}\frac{d_i}{n_i},
\end{equation}

\noindent where $d_i$ is the number of observed events at time $t_i$ and $n_i$ is the total number of patients at risk at time $t_i$. For each patient, instead of physical time $t_i$, we used the value of pseudotime computed along a particular trajectory.

For computing multivariate survival regression, we used the standard Cox model using the object 'CoxPHFitter' from the same package.

\section{Availability of source code and requirements}

\begin{itemize}
\item Clinical trajectories (ClinTrajan)
\item RRID identifier: SCR\_019018
\item bio.tools page: https://bio.tools/clintrajan
\item Project home page: \url{https://github.com/auranic/ClinTrajan}
\item Operating system(s): Platform independent
\item Programming language: Python 3.*
\item Other requirements: none
\item License: LGPL
\end{itemize}

\section{Availability of supporting data and materials}

The data set(s) supporting the results of this article is(are) available in the [repository name] repository, [cite unique persistent identifier]. 

\section{Supplementary Figures}

Supplementary Figure 1. Intrinsic dimensionality analysis of clinical datasets used in the study. The PCA-based estimation is defined here as the number of the eigenvalues of the covariance matrix exceeding $\lambda_0/C$, where $\lambda_0$ is the first (largest) eigenvalue and $C$ is the maximal conditional number of the covariance matrix after dimensionality reduction (here $C=10$). The computations were performed using the package \url{https://github.com/j-bac/scikit-dimension}, where one can find the complete definitions of the methods and the corresponding references.

\begin{figure*}[bt!] 
\centering
\includegraphics[width=\linewidth]{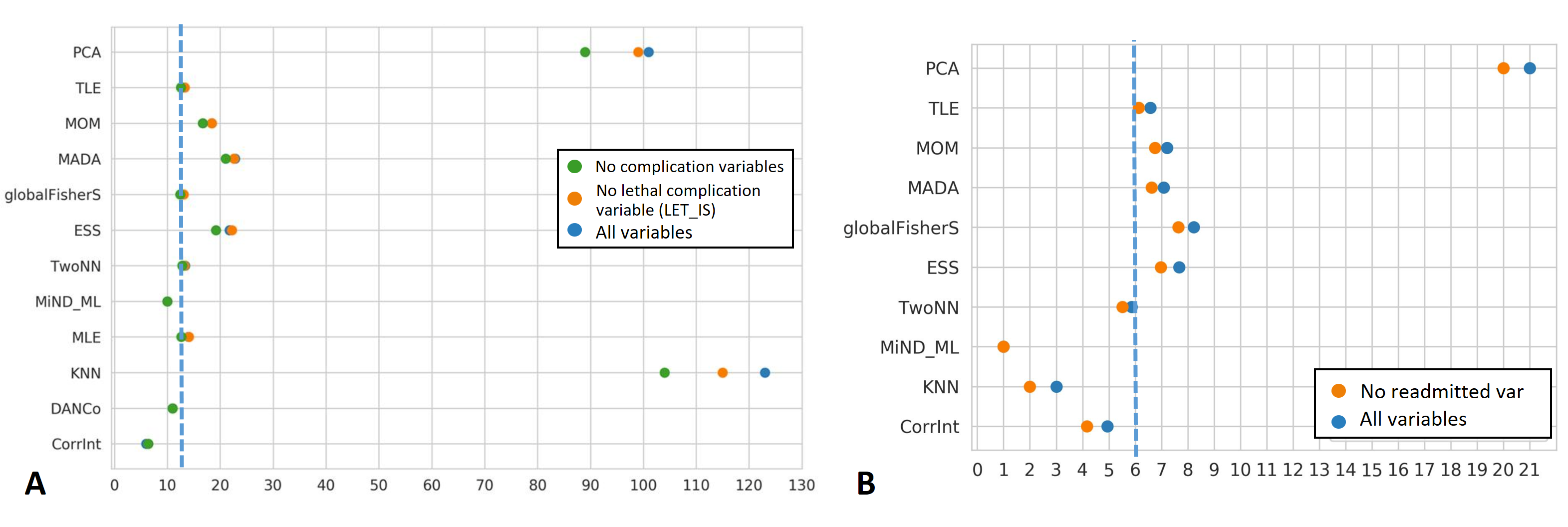}
\caption{Supplementary Figure 1. Intrinsic dimensionality analysis of clinical datasets used in the study. The PCA-based estimation is defined here as the number of the eigenvalues of the covariance matrix exceeding $\lambda_0/C$, where $\lambda_0$ is the first (largest) eigenvalue and $C$ is the maximal conditional number of the covariance matrix after dimensionality reduction (here $C=10$). The computations were performed using the package \url{https://github.com/j-bac/scikit-dimension}, where one can find the complete definitions of the methods and the corresponding references. }\label{fig:SupplementaryFigure1}
\end{figure*}

\section{Declarations}

\subsection{List of abbreviations}

\subsection{Consent for publication}

Not applicable.

\subsection{Competing Interests}

The author(s) declare that they have no competing interests

\subsection{Funding}

This work has been partially supported by the Ministry of Science and Higher Education of the Russian Federation (project No. 14.Y26.31.0022), by Agence Nationale de la Recherche in the program Investissements d’Avenir (project No. ANR-19-P3IA-0001; PRAIRIE 3IA Institute), by European Union’s Horizon 2020 program (grant No. 826121, iPC project), by the Association Science et Technologie, the Institut de Recherches Internationales Servier and the doctoral school Frontières de l’Innovation en Recherche et Education Programme Bettencourt.

\subsection{Author's Contributions}

A.Z., S.E.G., E.M.M. and A.N.G. designed the study. S.E.G., E.M.M. and Yu.O. prepared the myocardial infarction database, made it publicly available and advised on its quantification. A.Z., J.B., A.Ch., E.M.M., A.N.G., E.B. developed the methodology, based on application of elastic principal graphs and A.Z., J.B., A.Ch. implemented it in Python. J.B. packaged $ClinTrajan$. A.Z. and S.E.G. applied $ClinTrajan$ to the clinical datasets. S.E.G., A.Z., A.N.G. and Yu.O. participated in the interpretation of the results. A.Z. and S.E.G. drafted the manuscript. All authors participated in editing and finalizing the text.


\bibliography{paper-refs}

\begin{thebibliography}{47}
\providecommand{\natexlab}[1]{#1}
\providecommand{\url}[1]{\texttt{#1}}
\providecommand{\urlprefix}{}

\bibitem[{Jensen et~al.(2014)Jensen, Anders Boeck and Moseley, Pope L. and
  Oprea, Tudor I. and Elles{\o}e, Sabrina Gade and Eriksson, Robert and
  Schmock, Henriette and Jensen, Peter Bj{\o}dstrup and Jensen, Lars Juhl and
  Brunak, S{\o}ren}]{Jensen2014}
Jensen AB, Moseley PL, Oprea TI, Elles{\o}e SG, Eriksson R, Schmock H, et~al.
\newblock {Temporal disease trajectories condensed from population-wide
  registry data covering 6.2 million patients}.
\newblock Nature Communications 2014;5(1):1--10.

\bibitem[{Westergaard et~al.(2019)Westergaard, David and Moseley, Pope and
  S{\o}rup, Freja Karuna Hemmingsen and Baldi, Pierre and Brunak,
  S{\o}ren}]{Westergaard2019}
Westergaard D, Moseley P, S{\o}rup FKH, Baldi P, Brunak S.
\newblock {Population-wide analysis of differences in disease progression
  patterns in men and women}.
\newblock Nature Communications 2019;10(1):1--14.

\bibitem[{Moulis et~al.(2015)Moulis, G. and Lapeyre-Mestre, M. and Palmaro, A.
  and Pugnet, G. and Montastruc, J. L. and Sailler, L.}]{Moulis2015}
Moulis G, Lapeyre-Mestre M, Palmaro A, Pugnet G, Montastruc JL, Sailler L.
\newblock {French health insurance databases: What interest for medical
  research?}
\newblock Rev Med Interne 2015;36(6):411--417.

\bibitem[{Pinaire et~al.(2017)Pinaire, Jessica and Az{\'{e}},
  J{\'{e}}r{\^{o}}me and Bringay, Sandra and Landais, Paul}]{Pinaire2017}
Pinaire J, Az{\'{e}} J, Bringay S, Landais P.
\newblock {Patient healthcare trajectory. An essential monitoring tool: a
  systematic review}.
\newblock Health Information Science and Systems 2017;5(1):1.

\bibitem[{Albers et~al.(2014)Albers, D J and Tabak, E and Perotte, A and
  Hripcsak, George}]{Albers2014}
Albers DJ, Tabak E, Perotte A, Hripcsak G.
\newblock {Dynamical Phenotyping : Using Temporal Analysis of Clinically
  Collected Physiologic Data to Stratify Populations}.
\newblock PLoS ONE 2014;9(6):e96443.

\bibitem[{Ruderman(2017)Ruderman, Daniel}]{Ruderman2017}
Ruderman D.
\newblock {The emergence of dynamic phenotyping}.
\newblock Cell Biology and Toxicology 2017;33:507--509.

\bibitem[{Wang et~al.(2017)Wang, William and Zhu, Bijun and Wang,
  Xiangdong}]{Wang2017}
Wang W, Zhu B, Wang X.
\newblock {Dynamic phenotypes : illustrating a single-cell odyssey}.
\newblock Cell Biology and Toxicology 2017;33:423--427.

\bibitem[{Xu and Wunsch(2008)Xu, Rui and Wunsch, Donald C.}]{Xu2008}
Xu R, Wunsch DC.
\newblock {Clustering}.
\newblock IEEE Press, Piscataway, NJ; 2008.

\bibitem[{Jung and Wickrama(2008)Jung, Tony and Wickrama, K. A. S.}]{Jung2008}
Jung T, Wickrama KAS.
\newblock {An Introduction to Latent Class Growth Analysis and Growth Mixture
  Modeling}.
\newblock Social and Personality Psychology Compass 2008;2(1):302--317.

\bibitem[{Nagin and Odgers(2010)Nagin, Daniel S. and Odgers, Candice
  L.}]{Nagin2010}
Nagin DS, Odgers CL.
\newblock {Group-Based Trajectory Modeling in Clinical Research}.
\newblock Annual Review of Clinical Psychology 2010;(6):109--138.

\bibitem[{Rizopoulos(2011)Rizopoulos, Dimitris}]{Rizopoulos2011}
Rizopoulos D.
\newblock {Dynamic Predictions and Prospective Accuracy in Joint Models for
  Longitudinal and Time-to-Event Data}.
\newblock Biometrics 2011;67(3):819--829.
\newblock \urlprefix\url{https://www.jstor.org/stable/41242530}.

\bibitem[{Schulam et~al.(2015)Schulam, Peter and Wigley, Fredrick and Saria,
  Suchi}]{Schulam2015}
Schulam P, Wigley F, Saria S.
\newblock {Clustering longitudinal clinical marker trajectories from electronic
  health data: Applications to phenotyping and endotype discovery}.
\newblock In: Proceedings of the Twenty-Ninth AAAI Conference on Artificial
  Intelligence; 2015. p. 2956--2964.

\bibitem[{Schulam and Arora(2016)Schulam, Peter and Arora, Raman}]{Schulam2016}
Schulam P, Arora R.
\newblock {Disease trajectory maps}.
\newblock In: Proceedings of the Thirtieth Conference on Neural Information
  Processing Systems; 2016. .

\bibitem[{Banaee et~al.(2013)Banaee, Hadi and Ahmed, Mobyen Uddin and Loutfi,
  Amy}]{Banaee2013}
Banaee H, Ahmed MU, Loutfi A.
\newblock {Data mining for wearable sensors in health monitoring systems: A
  review of recent trends and challenges}.
\newblock {Sensors (Basel)} 2013;13(12):17472--17500.

\bibitem[{Chen et~al.(2019)Chen, Huidong and Albergante, Luca and Hsu, Jonathan
  Y. and Lareau, Caleb A. and {Lo Bosco}, Giosu{\`{e}} and Guan, Jihong and
  Zhou, Shuigeng and Gorban, Alexander N. and Bauer, Daniel E. and Aryee,
  Martin J. and Langenau, David M. and Zinovyev, Andrei and Buenrostro, Jason
  D. and Yuan, Guo Cheng and Pinello, Luca}]{Chen2019}
Chen H, Albergante L, Hsu JY, Lareau CA, {Lo Bosco} G, Guan J, et~al.
\newblock {Single-cell trajectories reconstruction, exploration and mapping of
  omics data with STREAM}.
\newblock Nature Communications 2019;10(1903).

\bibitem[{Saelens et~al.(2019)Saelens, Wouter and Cannoodt, Robrecht and
  Todorov, Helena and Saeys, Yvan}]{Saelens2019}
Saelens W, Cannoodt R, Todorov H, Saeys Y.
\newblock {A comparison of single-cell trajectory inference methods}.
\newblock Nature Biotechnology 2019;37(5):547--554.

\bibitem[{Gorban et~al.(2007)Gorban, A N and Sumner, N R and Zinovyev, A
  Y}]{Gorban2007Topological}
Gorban AN, Sumner NR, Zinovyev AY.
\newblock {Topological grammars for data approximation}.
\newblock Applied Mathematics Letters 2007;20(4):382 -- 386.
\newblock
  \urlprefix\url{http://www.sciencedirect.com/science/article/pii/S0893965906001856}.

\bibitem[{Gorban and Zinovyev(2008)Gorban, A. N. and Zinovyev, A.
  Y.}]{Gorban2008Principal}
Gorban AN, Zinovyev AY.
\newblock {Principal Graphs and Manifolds}.
\newblock In: Handbook of Research on Machine Learning Applications and Trends:
  Algorithms, Methods and Techniques (ed. E.Olivas) Information Science
  Reference, Hershey, PA;
  2008.\urlprefix\url{http://arxiv.org/abs/0809.0490{\%}0Ahttp://dx.doi.org/10.4018/978-1-60566-766-9}.

\bibitem[{Albergante et~al.(2020)Albergante, Luca and Mirkes, Evgeny and Bac,
  Jonathan and Chen, Huidong and Martin, Alexis and Faure, Louis and Barillot,
  Emmanuel and Pinello, Luca and Gorban, Alexander and Zinovyev,
  Andrei}]{Albergante2020}
Albergante L, Mirkes E, Bac J, Chen H, Martin A, Faure L, et~al.
\newblock {Robust and scalable learning of complex intrinsic dataset geometry
  via ElPiGraph}.
\newblock Entropy 2020;22(3):296.

\bibitem[{Parra et~al.(2019)Parra, R Gonzalo and Papadopoulos, Nikolaos and
  Ahumada-Arranz, Laura and Kholtei, Jakob El and Mottelson, Noah and
  Horokhovsky, Yehor and Treutlein, Barbara and Soeding, Johannes}]{Parra2019}
Parra RG, Papadopoulos N, Ahumada-Arranz L, Kholtei JE, Mottelson N,
  Horokhovsky Y, et~al.
\newblock {Reconstructing complex lineage trees from scRNA-seq data using
  MERLoT}.
\newblock Nucleic Acids Research 2019;47(17):8961--8974.
\newblock
  \urlprefix\url{https://academic.oup.com/nar/article/47/17/8961/5552070}.

\bibitem[{Marso et~al.(1999)Marso, Steven P and Griffin, Brian P. and Topol,
  Eric J.}]{Marso1999}
Marso SP, Griffin BP, Topol EJ, editors.
\newblock {Manual of Cardiovascular Medicine}.
\newblock Lippincott Williams {\&} Wilkins, Philadelphia, Pennsylvania, US;
  1999.

\bibitem[{Golovenkin et~al.(2020)Golovenkin, S. E. and Gorban, A.N. and Mirkes,
  Evgeny M. and Shulman, VA and Rossiev, D. A. and Shesternya, D.A. and
  Nikulina, S.Yu. and Orlova, Yu.V. and Dorrer, M.G.}]{Golovenkin2020}
Golovenkin SE, Gorban AN, Mirkes EM, Shulman V, Rossiev DA, Shesternya DA,
  et~al., {Myocardial infarction complications Database}.
\newblock Dataset; 2020.
\newblock \urlprefix\url{https://doi.org/10.25392/leicester.data.12045261.v2}.

\bibitem[{{Gorban} et~al.(1995)A. N. {Gorban} and D. A. {Rossiev} and E. V.
  {Butakova} and S. E. {Gilev} and S. E. {Golovenkin} and S. A. {Dogadin} and
  M. G. {Dorrer} and D. A. {Kochenov} and A. G. {Kopytov} and E. V.
  {Maslennikova} and G. V. {Matyushin} and Y. M. {Mirkes} and B. V. {Nazarov}
  and K. G. {Nozdrachev} and A. A. {Savchenko} and S. V. {Smirnova} and V. A.
  {Shulman} and V. I. {Zenkin}}]{Gorban1995MultiNeuron}
{Gorban} AN, {Rossiev} DA, {Butakova} EV, {Gilev} SE, {Golovenkin} SE,
  {Dogadin} SA, et~al.
\newblock Medical and Physiological Applications of MultiNeuron Neural
  Simulator.
\newblock In: International Neural Network Society Annual Meeting; Lawrence
  Erlbaum Associates, vol.~1; 1995. p. 170--175.
\newblock \urlprefix\url{https://arxiv.org/abs/q-bio/0411034}.

\bibitem[{Zinovyev(2001)Zinovyev, Andrei Yu.}]{Zinovyev2001}
Zinovyev AY.
\newblock {Visualization of Multidimensional data [in Russian]}.
\newblock Krasnoyarsk, Russia: Krasnoyarsk State Technical Universtity; 2001.

\bibitem[{Potluri et~al.(2016)Potluri, Rahul and Drozdov, Ignat and Carter, PR
  and Sarma, Jaydeep}]{Potluri2016}
Potluri R, Drozdov I, Carter P, Sarma J.
\newblock {Big data and cardiology: time for mass analytics?}
\newblock Eur Med J 2016;1(2):15--17.

\bibitem[{Strack et~al.(2014)Strack, Beata and Deshazo, Jonathan and Gennings,
  Chris and Olmo Ortiz, Juan Luis and Ventura, Sebastian and Cios, Krzysztof
  and Clore, John}]{Strack2014}
Strack B, Deshazo J, Gennings C, Olmo~Ortiz JL, Ventura S, Cios K, et~al.
\newblock Impact of HbA1c Measurement on Hospital Readmission Rates: Analysis
  of 70,000 Clinical Database Patient Records.
\newblock BioMed research international 2014;2014:781670.

\bibitem[{Gorban et~al.(2008)Gorban, Alexander N and Sumner, Neil R and
  Zinovyev, Andrei Y}]{Gorban2008Beyond}
Gorban AN, Sumner NR, Zinovyev AY.
\newblock {Beyond the concept of manifolds: Principal trees, metro maps, and
  elastic cubic complexes}.
\newblock In: Principal manifolds for data visualization and dimension
  reduction (eds. Gorban A.N, Kegl, B., Wunsch D., Zinovyev A.) Lecture Notes
  in Computational Science and Engineering, Springer; 2008.p. 219--237.

\bibitem[{Casacci and Pareto(2015)Casacci, Sara and Pareto,
  Adriano}]{Casacci2015}
Casacci S, Pareto A.
\newblock {Methods for quantifying ordinal variables: a comparative study}.
\newblock Quality and Quantity 2015;49(5):1859--1872.
\newblock \urlprefix\url{http://dx.doi.org/10.1007/s11135-014-0063-2}.

\bibitem[{Long(2018)Long, Andrew}]{Long2018}
Long A, {Using Machine Learning to Predict Hospital Readmission for Patients
  with Diabetes with Scikit-Learn}; 2018.
\newblock
  \urlprefix\url{https://towardsdatascience.com/predicting-hospital-readmission-for-patients-with-diabetes-using-scikit-learn-a2e359b15f0}.

\bibitem[{Tarpey and Flury(1996)Tarpey, T and Flury, B}]{Tarpey1996}
Tarpey T, Flury B.
\newblock {Self-Consistency: A Fundamental Concept in Statistics}.
\newblock Statistical Science 1996;11(3):229--243.

\bibitem[{T and W(1989)Hastie T and Stuetzle W}]{Hastie1989}
T H, W S.
\newblock Principal Curves.
\newblock Journal of the American Statistical Association
  1989;84(406):502--516.

\bibitem[{Whitwell et~al.(2020)Whitwell, Harry J. and Bacalini, Maria Giulia
  and Blyuss, Oleg and Chen, Shangbin and Garagnani, Paolo and Gordleeva, Susan
  Yu and Jalan, Sarika and Ivanchenko, Mikhail and Kanakov, Oleg and Kustikova,
  Valentina and Mari{\~{n}}o, Ines P. and Meyerov, Iosif and Ullner, Ekkehard
  and Franceschi, Claudio and Zaikin, Alexey}]{Whitwell2020}
Whitwell HJ, Bacalini MG, Blyuss O, Chen S, Garagnani P, Gordleeva SY, et~al.
\newblock {The Human Body as a Super Network: Digital Methods to Analyze the
  Propagation of Aging}.
\newblock Frontiers in Aging Neuroscience 2020;12:136.

\bibitem[{Zinovyev and Mirkes(2013)Zinovyev, Andrei and Mirkes,
  Evgeny}]{Zinovyev2013}
Zinovyev A, Mirkes E.
\newblock {Data complexity measured by principal graphs}.
\newblock Computers and Mathematics with Applications 2013;65(10):1471--1482.

\bibitem[{Setty et~al.(2019)Setty, Manu and Kiseliovas, Vaidotas and Levine,
  Jacob and Gayoso, Adam and Mazutis, Linas and Pe'er, Dana}]{Setty2019}
Setty M, Kiseliovas V, Levine J, Gayoso A, Mazutis L, Pe'er D.
\newblock {Characterization of cell fate probabilities in single-cell data with
  Palantir}.
\newblock Nature Biotechnology 2019;37:451--460.

\bibitem[{Saria(2018)Saria, Suchi}]{Saria2018}
Saria S.
\newblock {Individualized sepsis treatment using reinforcement learning}.
\newblock Nature Medicine 2018;24:1641--1642.

\bibitem[{CT et~al.(2012)Chen CT and Hung HM and Hsiao CF}]{Chen2012Design}
CT C, HM H, CF H.
\newblock Design and evaluation of multiregional trials with heterogeneous
  treatment effect across regions.
\newblock {Journal of Biopharmaceutical Statistics} 2012;22(5):1037--1050.

\bibitem[{Young(1981)Young, Forrest W.}]{Young1981}
Young FW.
\newblock {Quantitative analysis of qualitative data}.
\newblock Psychometrika 1981;46:357–388.

\bibitem[{Linting and {Van Der Kooij}(2012)Linting, Mari{\"{e}}lle and {Van Der
  Kooij}, Anita}]{Linting2012}
Linting M, {Van Der Kooij} A.
\newblock {Nonlinear principal components analysis with CATPCA: A tutorial}.
\newblock Journal of Personality Assessment 2012;94(1):12--25.

\bibitem[{Fehrman et~al.(2019)Fehrman, E and Egan, V and Gorban, Alexander N.
  and Levesley, Jeremy and Mirkes, E.M. and Muhammad, AK}]{Fehrman2019}
Fehrman E, Egan V, Gorban AN, Levesley J, Mirkes EM, Muhammad A.
\newblock {Personality Traits and Drug Consumption: a story told by data}.
\newblock Springer Berlin / Heidelberg; 2019.

\bibitem[{Mirkes et~al.(2016)E.M. Mirkes and T.J. Coats and J. Levesley and
  A.N. Gorban}]{Mirkes2016}
Mirkes EM, Coats TJ, Levesley J, Gorban AN.
\newblock Handling missing data in large healthcare dataset: A case study of
  unknown trauma outcomes.
\newblock Computers in Biology and Medicine 2016;75:203--216.
\newblock
  \urlprefix\url{http://www.sciencedirect.com/science/article/pii/S0010482516301421}.

\bibitem[{Dergachev et~al.(2001)Dergachev, V A and Gorban, A N and Rossiev, A A
  and Karimova, L M and Kuandykov, E B and Makarenko, N G and Steier,
  P}]{Dergachev2001}
Dergachev VA, Gorban AN, Rossiev AA, Karimova LM, Kuandykov EB, Makarenko NG,
  et~al.
\newblock The Filling of Gaps in Geophysical Time Series by Artificial Neural
  Networks.
\newblock Radiocarbon 2001;43(2A):365--371.

\bibitem[{Albergante et~al.(2019)Albergante, Luca and Bac, Jonathan and
  Zinovyev, Andrei}]{Albergante2019}
Albergante L, Bac J, Zinovyev A.
\newblock {Estimating the effective dimension of large biological datasets
  using Fisher separability analysis}.
\newblock In: Proceedings of the International Joint Conference on Neural
  Networks; 2019. .

\bibitem[{Gorban et~al.(2008)Gorban, A and K{\'{e}}gl, Bal{\^{a}}zs and Wunch,
  D and Zinovyev, Andrei}]{Gorban2008book}
Gorban A, K{\'{e}}gl B, Wunch D, Zinovyev A, editors.
\newblock {Principal Manifolds for Data Visualisation and Dimension Reduction}.
\newblock Lecture notes in Computational Science and Engineering, Springer,
  Berlin; 2008.

\bibitem[{Gorban and Zinovyev(2010)Gorban, Alexander N and Zinovyev,
  Andrei}]{Gorban2010Principal}
Gorban AN, Zinovyev A.
\newblock {Principal manifolds and graphs in practice: from molecular biology
  to dynamical systems.}
\newblock International Journal of Neural Systems 2010;20(3):219--232.
\newblock \urlprefix\url{http://arxiv.org/abs/1001.1122}.

\bibitem[{Gorban et~al.(2017)Gorban, Alexander N. and Mirkes, Evgeny and
  Zinovyev, Andrei Yu.}]{Gorban2017Robust}
Gorban AN, Mirkes E, Zinovyev AY.
\newblock {Robust principal graphs for data approximation}.
\newblock Archives of Data Science 2017;2(1):1:16.

\bibitem[{Kamada et~al.(1989)Kamada, Tomihisa and Kawai, Satoru and
  others}]{kamada1989algorithm}
Kamada T, Kawai S, et~al.
\newblock An algorithm for drawing general undirected graphs.
\newblock Information processing letters 1989;31(1):7--15.

\bibitem[{Nelson(1969)Nelson, W}]{Nelson1969}
Nelson W.
\newblock {Hazard plotting for incomplete failure data}.
\newblock Journal of Quality Technology 1969;1:27--52.

\end{thebibliography}

\end{document}